\documentclass[useAMS,usenatbib,usegraphicx]{mn2e}
\usepackage{aas_macros}
\usepackage[british]{babel}
\usepackage{multirow}
\bibpunct{(}{)}{,}{a}{}{,} 
\newcommand{\lt}{<}

\newcommand{\hr}{HR}
\newcommand{\pms}{pre-MS}
\newcommand{\zams}{ZAMS}
\newcommand{\ms}{MS}

\newcommand{\prosecco}{\textsc{prosecco}}

\newcommand{\dv}{LA}
\newcommand{\std}{STD}
\newcommand{\psa}{PLA}

\newcommand{\gc}{GC}
\newcommand{\gcs}{GCs}
\newcommand{\msun}{M$_{\sun}$}
\newcommand{\rsun}{R$_{\sun}$}

\title[Accreting PMS models in globular clusters]{Accreting pre-main sequence models and abundance anomalies in globular clusters}
\author[E. Tognelli, P. G. Prada Moroni, \& S. Degl'Innocenti]
{E. Tognelli$^{1,2}$\thanks{e-mail: emanuele.tognelli$@$for.unipi.it}, P.G. Prada Moroni$^{2,3}$\thanks{e-mail: pier.giorgio.prada.moroni$@$unipi.it}, 
S. Degl'Innocenti$^{2,3}$\\
$^{1}$University of Roma Tor Vergata, Department of Physics, Via della Ricerca Scientifica 1, I-00133 Roma, Italy\\
$^{2}$INFN, Section of Pisa, Largo Bruno Pontecorvo 3, I-56127 Pisa, Italy\\
$^{3}$University of Pisa, Department of Physics `E.Fermi', Largo Bruno Pontecorvo 3, I-56127 Pisa, Italy
}
\begin{document}
\date{Accepted 2015 September 28.  Received 2015 September 8; in original form 2015 June 30}
\pagerange{\pageref{firstpage}--\pageref{lastpage}} \pubyear{2015}
\maketitle
\label{firstpage}
\begin{abstract}
We investigated the possibility of producing helium enhanced stars in globular clusters by accreting polluted matter during the pre-main sequence phase. We followed the evolution of two different classes of pre-main sequence accreting models, one which neglects and the other that takes into account the protostellar evolution. 

We analysed the dependence of the final central helium abundance, of the tracks position in the \hr{} diagram and of the surface lithium abundance evolution on the age at which the accretion of polluted material begins and on the main physical parameters that govern the protostellar evolution. The later is the beginning of the late accretion and the lower are both the central helium and the surface lithium abundances at the end of the accretion phase and in \zams{} (Zero Age Main Sequence). In order to produce a relevant increase of the central helium content the accretion of polluted matter should start at ages lower than 1 Myr. The inclusion of the protostellar evolution has a strong impact on the \zams{} models too. The adoption of a very low seed mass (i.e. 0.001~\msun) results in models with the lowest central helium and surface lithium abundances. The higher is the accretion rate and the lower is the final helium content in the core and the residual surface lithium. In the worst case -- i.e. seed mass 0.001~\msun{} and accretion rate $\ge 10^{-5}$~\msun{} yr$^{-1}$ -- the central helium is not increased at all and the surface lithium is fully depleted in the first few million years.
\end{abstract}
\begin{keywords}
stars: abundances -- stars: evolution -- stars: formation -- stars: general -- stars: low-mass -- stars: pre-main sequence
\end{keywords}
\maketitle
\section{Introduction}
\citet{bastian13} recently proposed a new scenario for the formation of multi-populations in globular clusters (\gcs) which does not require multiple episodes of star formation. Such a scenario has the advantage of alleviating the mass budget problems that affects the other suggested mechanism that is, the necessity of having \gcs{} with initial mass much higher -- by a factor of 10--100 -- than the current one \citep{renzini08}. Modifying an earlier idea by \citet{dantona83} and \citet{thoul02}, they hypothesized that the stars that show chemical anomalies, i.e. oxygen depletion and sodium enhancement, are actually born in the same burst which produced the stars with standard chemical patterns rather than in a second episode of stellar formation. The observed chemical anomalies would be the result of an accretion of polluted matter on the disc surrounding pre-main sequence (\pms) stars. Since \pms{} stars are initially fully convective, polluted matter would be naturally mixed, and thus diluted, in the whole stellar structure. This is a key point because chemical anomalies are found both in turn off and red giant branch. 

\citet{bastian13} assumed that the polluters are massive interacting binaries of 15--20~\msun{} that provide nuclear-processed matter after about 2--3 Myr. They also assumed that the accretion disc survive for about 10--20 Myr, enough to build the observed second generation stars. However, \citet{mamajek09} and \citet{ribas14} showed that the fraction of stars with discs exponentially decreases as the age increases, with a characteristic time-scale of about 2--5 Myr. Consequently, if the efficient phase of accretion occurs on a time-scale typical of the disc lifetime \citep[2--5 Myr instead of the 10--20 Myr assumed by][]{bastian13}, this might significantly reduce the amount of polluted matter available, thus weakening the capability of the accretion model scenario to produce second generation stars without requiring \gcs{} with very high initial mass.

\citet{cassisi14} and \citet{salaris14} investigated the impact of accretion of polluted matter during the \pms{} phase on the predicted abundances of central helium and surface lithium. Regarding the central helium abundance, \citet{cassisi14} showed that it is 
possible to produce a population of helium-rich stars with a small dispersion in central helium abundances as it seems required by observations of some \gc, as NGC~2808 \citep{piotto07}. On the other hand, \citet{salaris14} showed that the accretion scenario is not able to provide polluted stars with a sizeable residual surface lithium abundance, unless the accretion is confined within the first 2--3~Myr, thus significantly reducing the advantages of this scenario in alleviating the mass budget problem. However, in these studies the accretion evolution is not actually followed and the final abundances of helium and lithium are computed by assuming that the accreted matter is completely mixed with the original one in fully convective stars. Moreover, it is assumed that the accretion process would not modify the structure of \pms{} stars. To this regard, \citet{dantona14} proved that such an assumption is not always fulfilled, as the accretion modifies the stellar structure during the \pms{} phase in such a way that masses larger than 0.4~\msun{} develop a radiative core before the end of the accretion, hence preventing that polluted matter could reach the stellar centre. This result further reduces the ability of the proposed scenario to produce high central helium enhanced stars without the need of an initial very massive cluster.

\citet{dantona14} followed the accretion of polluted matter on stars of initial mass $0.1\,$\msun$\le M_0\le 0.5\,$\msun up to a final mass $0.2\,$\msun$\le M_\rmn{fin}\le 0.8\,$\msun. Regardless the value of the initial mass $M_0$, all stars start the accretion at the same time, i.e. at the beginning of their \pms{} phase along the Hayashi track. 

Another crucial issue of the accretion models concerns the amount of residual surface lithium in turn-off and subgiant stars. \citet{salaris14} and \citet{dantona14} showed that the inclusion of accretion of lithium-free material leads to a drastic reduction of the surface lithium content in turn-off and subgiant second generation stars. Such an occurrence conflicts with the observations of a relatively large amount of residual lithium in turn-off and subgiant stars of both first and second generation \citep[see e.g.][]{dorazi10,monaco12,dorazi14,dorazi15} which is similar to that measured in Population II field stars \citep[see e.g.][]{pasquini94,pasquini96,bonifacio02,mucciarelli11}.

We aim to further explore the \citet{bastian13} scenario by investigating two aspects not yet studied. First, the effect of varying the starting age ($t_{0,\rmn{l.acc}}$) of the accretion process, as the polluting matter is not yet available at the beginning of \pms{} evolution. According to \citet{bastian13} massive interacting binaries are able to provide a sizeable amount of polluted matter not before 2--3~Myr. The later is the beginning of the accretion, i.e. the larger $t_{0,\rmn{l.acc}}$, and the more advanced is the evolutionary phase of a \pms{} star along the Hayashi track, thus the shorter is the time interval of accretion on to a fully convective star. Moreover, since stars of different mass evolve at different evolutionary speeds, the impact of changing $t_{0,\rmn{l.acc}}$ depends on mass. At a given $t_{0,\rmn{l.acc}}$, the more massive is a star and in more advanced phase accretion begins, thus the shorter is the duration of accretion on fully convective structure. In both cases, the earlier formation of the radiative core leads to a lower central helium enhancement. In Section \ref{sec:dv}, we will show that models that start accreting polluted matter at 2--4~Myr rather than at zero age have a systematically lower helium content both at the end of the late accretion phase and close to the \zams. 

Another aspect never studied before in this contest is the effect of protostellar accretion, i.e. the accretion on the first hydrostatic core which actually builds the star. \citet{dantona14} assumed that the accretion of polluted matter occurs on the Hayashi track of standard \pms{} stars, completely neglecting the previous protostellar evolution. In Section \ref{sec:psa} we will describe the evolution of models which undergo two different accretion episodes. The first accretion phase -- i.e. the protostellar accretion -- starts on a very low mass seed ($M_\rmn{seed} =$~0.001--0.05~\msun) and accretes non-polluted matter until the mass reaches the value $M_0$, while the second accretion phase -- i.e. the late accretion -- begins at $t_{0,\rmn{l.acc}}$ on $M_0$ and accretes polluted matter up to a final mass $M_\rmn{fin}$. 

We discuss the effect on the final central helium enhancement and the surface lithium abundance of changing the seed mass $M_\rmn{seed}$, the protostellar mass accretion rate $\dot{m}$ and the initial deuterium abundance $X_\rmn{d}$, within the current uncertainty range, during the protostellar evolution phase.

\section{Treatment of accretion in evolutionary models}
\label{sec:reference}
Stellar models have been computed adopting the most recent version of the Pisa stellar evolutionary code \citep[\prosecco{} see e.g.,][ and references therein for more details]{tognelli15a,tognelli15b} derived from the well tested \textsc{franec} one \citep[see e.g.][]{deglinnocenti08,tognelli11,tognelli12,dellomodarme12}. We recently modified our code to account for mass accretion from the boundary layer of a thin disc adopting the same formalism described in \citet{siess97}. The accretion is supposed to interest a very small part of the stellar surface so that the outer boundary conditions can be considered with a good approximation to be the same of a non-accreting model \citep[see e.g.][]{hartmann97}.

In this simplified case the parameters governing the accretion process are the mass accretion rate ($\dot{m}$), the mass fraction abundance of the $j^\rmn{th}$-element in the accreted matter ($X_{j,\rmn{acc}}$), and the fraction of in-falling energy ($\alpha_\rmn{acc}$) gained by the star. In the following we assumed a \emph{cold accretion model}, which means that the in-falling matter reaches the stellar surface with an internal energy content negligible with respect to the stellar one ($\alpha_\rmn{acc}=0$). 

We assumed convective mixing time-scale to be shorter than accretion one, hence the accreted matter is instantaneously and homogeneously distributed in the convective envelope. 

To analyse the effect of a \emph{late accretion phase} with matter enriched with polluted helium, we computed two classes of models. 

As a first step we used an approach similar to that presented in \citet{dantona14} where only the late accretion is actually followed (hereafter \dv{} models). Late accretion is supposed to occur on an already formed and homogeneous star evolved -- at constant mass -- from a cold initial model on the Hayashi track, with an arbitrary large radius and luminosity which neglect the previous protostellar evolution. The star has an \emph{initial mass} $M_0$ and an initial helium and metal abundance $Y_\rmn{ini}$ and $Z_\rmn{ini}$. Then, during the late accretion of polluted matter, both the mass and the chemical composition are modified, until a \emph{final mass} $M_\rmn{fin}$ is reached. 

We selected two $M_0$ and $M_\rmn{fin}$ values among those analysed by \citet{dantona14}, namely $M_0$~=~0.2 and 0.4~\msun, and $M_\rmn{fin}~=~0.5$ and 0.7~\msun. The chemical composition of the accreted matter has the same global metallicity ($Z_\rmn{ini}$) and metal distribution of the initial model but a significantly enhanced helium abundance ($Y_\rmn{l.acc}$) that we fixed to $Y_\rmn{l.acc}~=~0.440$ \citep{cassisi14}. The accreted matter is fully depleted of deuterium ($X_\rmn{d,l.acc}~=~0$) and lithium (and in general of light elements, i.e. boron and  beryllium), as the gas ejected by the polluters is expected to be depleted of such elements \citep{decressin07,demink09,frischkenecht10,brott11,salaris14}.

The mass accretion rate during the late accretion phase ($\dot{m}_\rmn{l.acc}$) has been chosen in order to produce the requested final mass ($M_\rmn{fin}$) in a fixed interval of time ($\Delta t_\rmn{l.acc}$), i.e.,
\begin{equation}
\dot{m}_\rmn{l.acc} \equiv \frac{M_\rmn{fin} - M_0}{\Delta t_\rmn{l.acc}}
\end{equation}
The late accretion phase is supposed to begin at a stellar age $t_{0,\rmn{l.acc}}$. Unless not explicitly stated, we adopted as reference the following values for $\Delta t_\rmn{l.acc}$ and $t_\rmn{0,l.acc}$, namely $\Delta t_\rmn{l.acc}~=~10$~Myr and $t_\rmn{0,l.acc}~=~2$~Myr (see Section \ref{sec:dv}).

As a further step, we included in the stellar model also the \emph{first accretion phase}, which is the accretion from the protostellar (hydrostatic) core. In this second case, both the protostellar and the late accretion phases are actually followed (hereafter \emph{protostellar-late accretion} models, \psa). We started the accretion from an initial hydrostatic core with a mass $M_\rmn{seed}$, the typical protostellar second Larson core \citep[see e.g.][and references therein]{larson69,baraffe12}. The accretion proceeds with the initial chemical composition ($Y_\rmn{ini}$, $Z_\rmn{ini}$, $X_\rmn{d,ini}$) at a constant accretion rate $\dot{m}$ until the star reaches the requested mass ($M_0$), when the accretion is switched off. Then, starting from $t_\rmn{0,l.acc}$ the star undertakes also the late accretion phase (with $Y_\rmn{l.acc}$, and $\dot{m}_\rmn{l.acc}$) until the final $M_\rmn{fin}$ is reached. Notice that $M_0$ is the same value used to start the late accretion in the \dv{} models previously described. Such a very simplified scheme is devised to make easier the comparison with the \dv{} models keeping fixed $M_0$ and $M_\rmn{fin}$. We set the reference values of the first accretion phase parameters as it follows: $M_\rmn{seed}~=~10^{-3}$~\msun, $R_\rmn{seed} =~1$~\rsun, and $\dot{m}~=~10^{-5}$~\msun~yr$^{-1}$. We will analyse the effect of adopting different values for these parameters in Section \ref{sec:psa}.

All the models presented in this work have been computed using an initial helium and metal content ($Y_\rmn{ini}$,~$Z_\rmn{ini}$)~=~(0.252,~0.002) and [$\alpha$/Fe]$~=~+0.3$ (i.e. [Fe/H]$=-1.0$), similar to what adopted by \citet{dantona14} and representative of NGC~2808 \citep[see e.g.][]{piotto07}. The initial deuterium abundance has been set to $X_\rmn{d,ini}~=~4\times 10^{-5}$ \citep[][]{pettini08,cooke14}.

Stellar modes have been computed adopting a mixing length parameter value of $\alpha_\rmn{ML}~=~1.00$, which is more suitable for the evolution of \pms{} stars \citep[see e.g.,][and references therein]{gennaro12,tognelli12}. However, to the sake of completeness we also computed accreting and non-accreting models with a solar calibrated mixing length parameter (i.e. $\alpha_\rmn{ML}~=~1.74$). A more efficient super adiabatic convection does not qualitatively change the impact of accretion on the central helium abundance evolution but it significantly enhances the \pms{} surface lithium depletion \citep[see e.g.][]{tognelli12}.

\section{Pre-MS models with the sole late accretion}
\label{sec:dv}
\begin{figure*}
	\centering
	\includegraphics[width=0.8\linewidth]{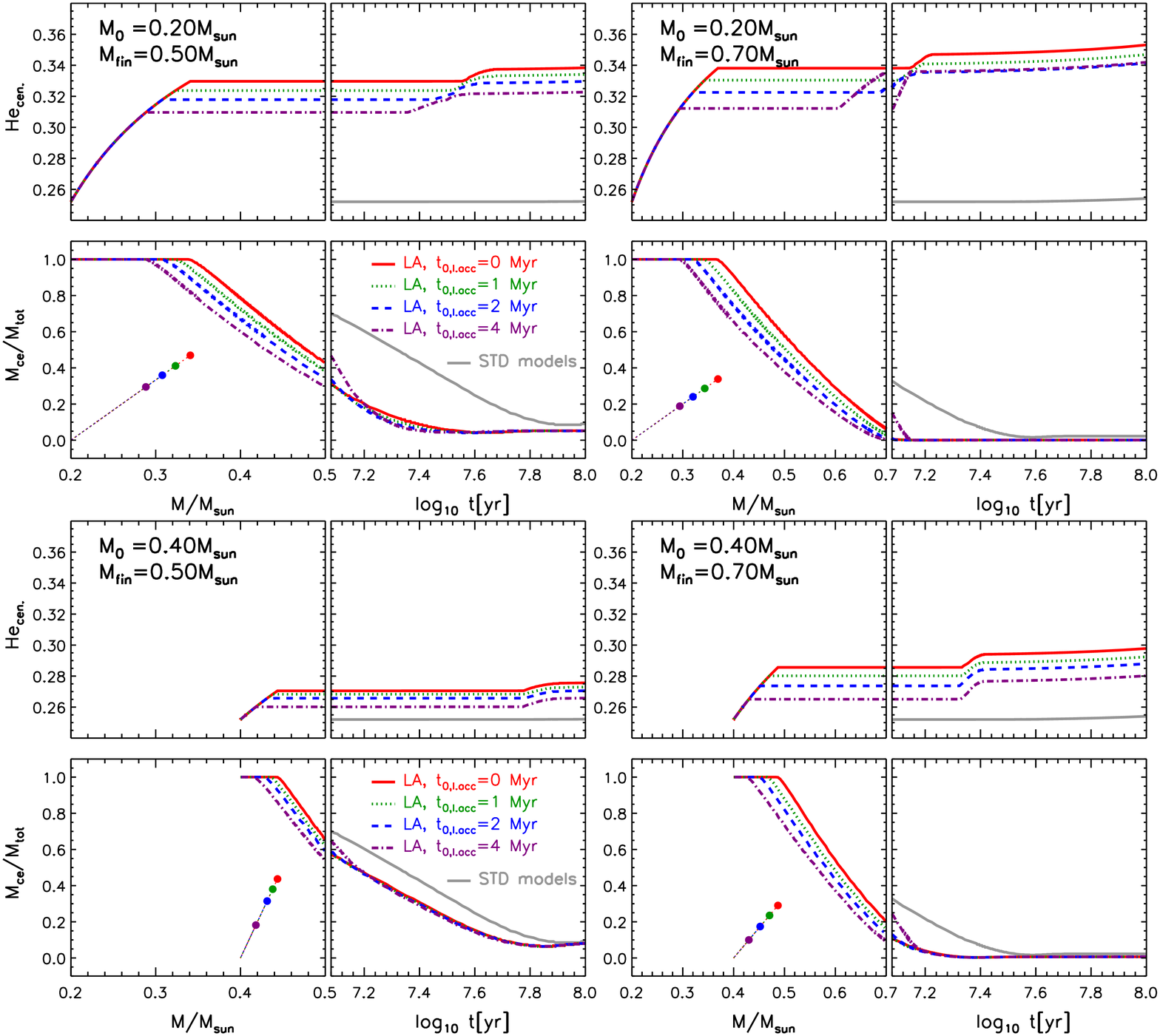}
	\caption{Evolution of the central helium mass fraction abundance (He$_\rmn{cen.}$) and of the fraction of stellar mass inside the convective envelope ($M_\rmn{ce}$/$M_\rmn{tot}$) as a function of both the total mass (during the accretion phase) and of the stellar age (for $t\ge12$~Myr) for the late accretion models (\dv). Models are computed adopting the four labelled values of the beginning of the late accretion starting ages ($t_\rmn{0,l.acc}$). Thin dotted lines in $M_\rmn{ce}$/$M_\rmn{tot}$ panel show the fraction of mass accreted during the fully convective phase (i.e. $M_\rmn{acc}^\rmn{fully conv.}/(M_\rmn{fin}-M_0)$); the end of such a phase is marked by a filled circle. Standard models with $M~=~M_\rmn{fin}$, $Y~=~Y_\rmn{ini}$ and $Z~=~Z_\rmn{ini}$ are also shown as a grey thick lines. Top panels: models with initial mass $M_0~=~0.2$~\msun{} and two values of the final mass, namely $M_\rmn{fin}~=~0.5$ \msun{} (left-hand panels) and $M_\rmn{fin}~=~0.7$~\msun{} (right-hand panels). Bottom panels: the same as the top panels but for $M_0~=~0.4$~\msun.}
	\label{fig:acc_dav_t0acc}
\end{figure*}
\begin{figure*}
	\centering
	\includegraphics[width=0.8\linewidth]{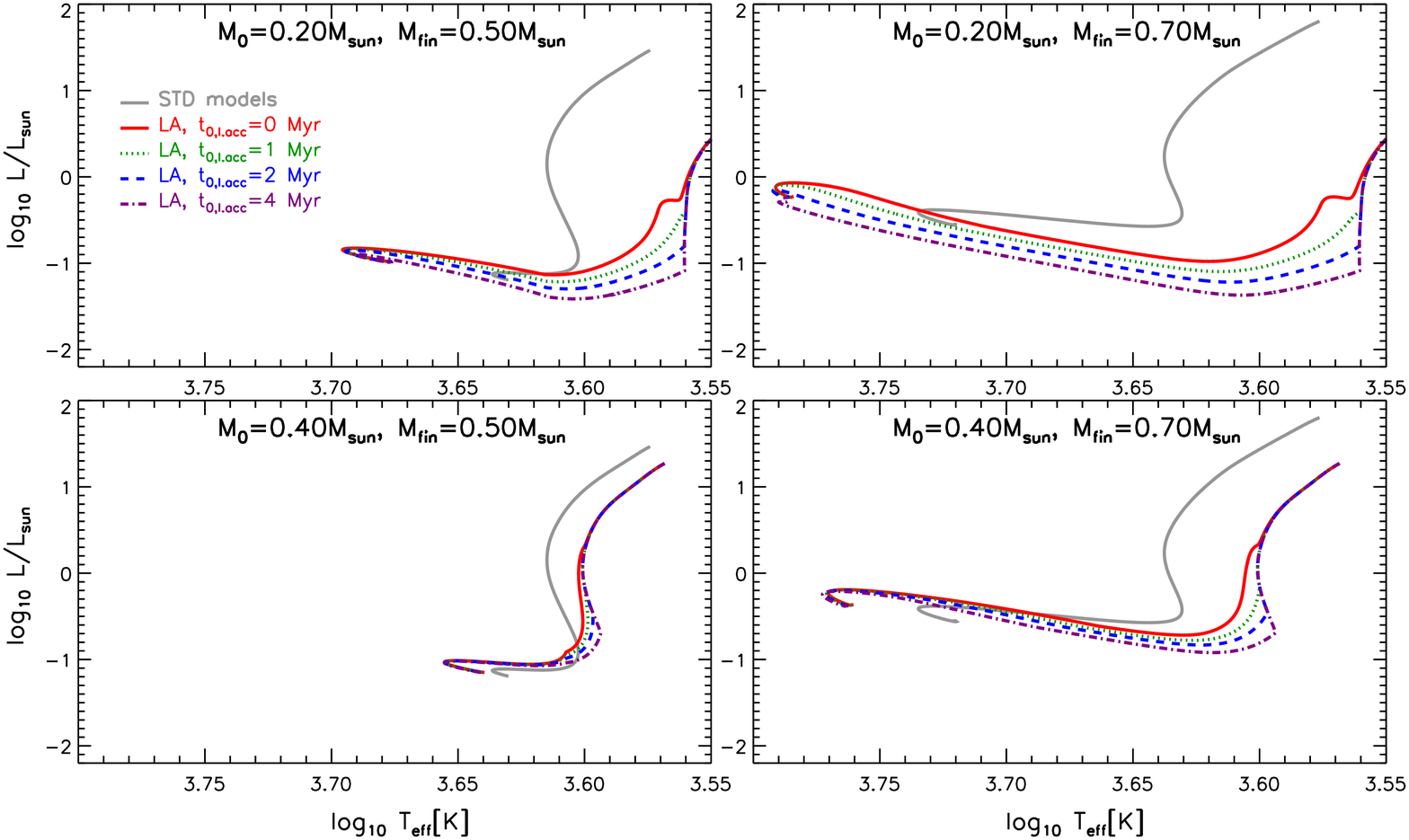}
	\caption{\hr{} diagrams for the late accretion models (\dv) shown in Fig. \ref{fig:acc_dav_t0acc}. Top panels: $M_0 = 0.2$ \msun{} with $M_\rmn{fin}~=~0.5$~\msun{} (left-hand panel) and $M_\rmn{fin}~=~0.7$~\msun{} (right-hand panel). The corresponding standard models are also shown (thick grey line). Bottom panel: the same as the top panel but for $M_0~=~0.4$~\msun.}
	\label{fig:acc_dav_t0acc_hr}
\end{figure*}
\begin{figure*}
	\centering
	\includegraphics[width=0.8\linewidth]{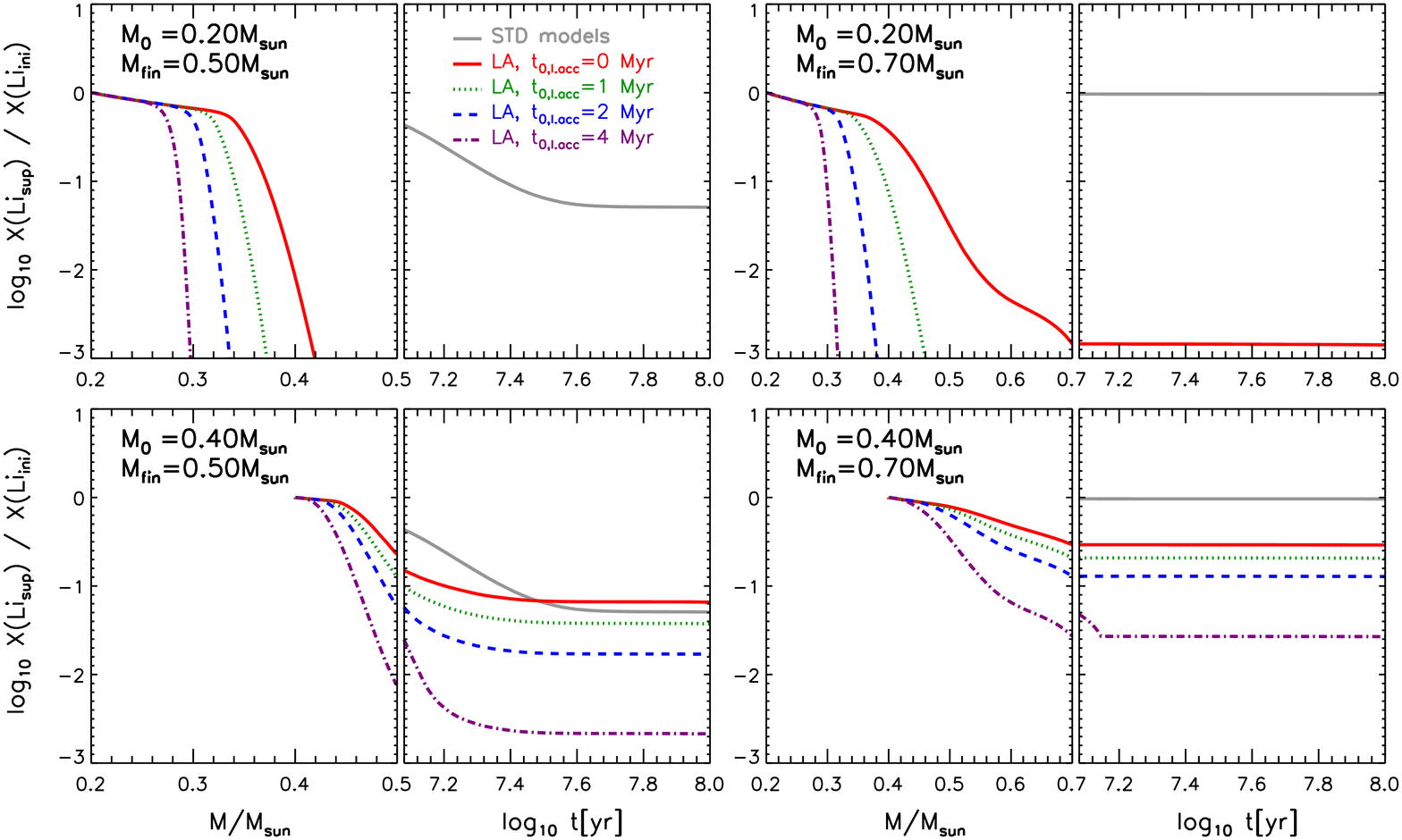}
	\caption{Surface $^7$Li abundance for the late accretion models (\dv) shown in Fig. \ref{fig:acc_dav_t0acc} as a function of both the total mass (during the accretion phase) and of the stellar age (for $t\ge12$~Myr). Top panels: $M_0~=~0.2$~\msun{} with $M_\rmn{fin}~=~0.5$~\msun{} (left-hand panel) and $M_\rmn{fin}~=~0.7$~\msun{} (right-hand panel). The corresponding standard models are also shown (thick grey line). Bottom panel: the same as the top panel but for $M_0~=~0.4$~\msun.}
	\label{fig:acc_dav_t0acc_li}
\end{figure*}

We explored the effect of a late mass accretion on standard \pms{} models starting their evolution on the Hayashi track at an initial radius and luminosity which do not take into account the protostellar phase. To compute the evolution of this class of models (\dv{}, only late accretion) we adopted the same choices of \citet{dantona14} with the exception of the starting age $t_\rmn{0,l.acc}$ of the late accretion, that was allowed to vary. This is an important difference to discuss, since  \citet{bastian13} showed that no polluters are available before $t~\sim~2$~Myr. 

The adoption of different $t_\rmn{0,l.acc}$ (at a fixed $M_0$) forces the accretion to begin in different evolutionary phases. In particular, at increasing $t_\rmn{0,l.acc}$ the accretion starts progressively closer to the first model that would develop a radiative core, with a consequent reduction of the amount of helium-rich material actually deposed in the stellar centre. On the other hand, at a fixed $t_\rmn{0,l.acc}$, the more massive is the initial mass $M_0$ and in a more advanced phase is the star at the beginning of the accretion, because the evolutionary speed increases with mass. Thus, the amount of matter accreted into the core (at a fixed $t_\rmn{0,l.acc}$) is reduced if $M_0$ increases. 

Fig. \ref{fig:acc_dav_t0acc} shows the central helium abundance and the mass of convective envelope as a function of the total mass during the accretion phase until $M_\rmn{fin}$ is reached and then as a function of the stellar age for $t \ge 12$~Myr, for \dv{} models. We adopted different values of $t_\rmn{0,l.acc}$, namely, $t_\rmn{0,l.acc}=0,\,1,\, 2,$ and 4 Myr, for initial masses $M_0~=~0.2$~\msun{} (upper panels) and 0.4~\msun{} (bottom panels) and final masses $M_\rmn{fin}~=~0.5$ and 0.7~\msun. The set with $t_\rmn{0,l.acc}~=~0$ has been computed to reproduce the same configuration adopted by \citet{dantona14}. Standard models without any accretion episode (hereafter \std) for the same final mass and with $Y~=~Y_\rmn{ini}$(=0.252) and $Z~=~Z_\rmn{ini}$(=0.002) are also shown for comparison. The mass fractional abundance of central and surface helium at the end of the late accretion phase and at an age of $10^8$ yr (which roughly corresponds to the \zams{} location) along with $\log T_\rmn{eff}$ and $\log L/\rmn{L}_{\sun}$ at $10^8$~yr for the quoted models are listed in Table \ref{tab:la_models}.
 
In all the presented cases, the initial part of the late accretion occurs on a fully convective star, thus leading to an increase of the central helium abundance. However, as already shown by \citet{dantona14} in the case $t_\rmn{0,l.acc}~=~0$, more massive models develop a radiative core before the accretion is finished, thus preventing the further increase of central helium abundance.

The change of the age $t_\rmn{0,l.acc}$ at which the late accretion starts has a significant effect on the duration of the fully convective phase, consequently on the final amount of helium deposed in the stellar centre. As shown in Fig. \ref{fig:acc_dav_t0acc}, the total mass value at which a radiative core forms decrease as $t_\rmn{0,l.acc}$ increases, passing from about 0.34~\msun{}  for $t_\rmn{0,l.acc}=0$ to 0.29~\msun{} for $t_\rmn{0,l.acc}=4$~Myr in the case of $M_0=0.2$~\msun, and from 0.45~\msun{} to 0.42~\msun{} in the $M_0=0.4$~\msun{} for $t_\rmn{0,l.acc}=0$ and $t_\rmn{0,l.acc}=4$~Myr, respectively. Notice that, since the accretion rate is constant, this means that the temporal duration of the fully convective phase is reducing. It is clear that at a fixed $M_0$ the use of a larger $t_\rmn{0,l.acc}$ leads to a reduction of the amount of polluted matter accreted during the fully convective phase and consequently to a lower central helium content.

The models which start accreting polluted matter at the beginning of \pms{} evolution (i.e. $t_\rmn{0,l.acc}=0$) reach consequently the highest central helium abundance at the end of the accretion phase. This case corresponds to the scenario first explored by \citet{dantona14}, whose results are in good agreement with ours. However, as already discussed, $t_\rmn{0,l.acc}~\la~2$ Myr seems to be an unlikely scenario, while the adoption of $t_\rmn{0,l.acc}~\ge~2$ Myr is more consistent with the \citet{bastian13} model. From the figure it is clear that if $t_\rmn{0,l.acc}~\ge~2$ Myr is used, then the accreted central helium is systematically lower than \citet{dantona14} results. 

Fig. \ref{fig:acc_dav_t0acc} also shows an interesting behaviour never described before. In all the models it is clearly visible an increase of the central helium abundance at the end or close to the end of the accretion phase. Such a behaviour is the result of the chemical profile left by the accretion inside the radiative region (after the formation of the radiative core) and the following growth of a temporary convective core triggered by the onset of the $^3$He($^3$He,~2p)$^4$He nuclear reaction. Indeed, during the accretion phase, once the radiative core forms and starts to grow, the base of the region of the star actually polluted by the accreted matter progressively withdraws outwards. As a result, the helium abundance profile in the radiative core increases from the centre towards the convective envelope. Depending on the stellar mass and the helium profile left inside the radiative region, the temporary convective core might extend enough to significantly increase the central helium abundance, as in the cases of $M_\rmn{fin}~=~0.7$~\msun. 

Notice that the models which start accretion in advanced \pms{} phase (i.e. $t_\rmn{0,l.acc}~>~0$) also close to the \zams{} (i.e. $t\sim 10^8$~yr) -- after the formation of the convective core -- show always a central helium abundance lower than that predicted by models which start accretion just at the beginning of their \pms{} (i.e. $t_\rmn{0,l.acc}~=~0$). Such a result further weakens the ability of the accretion scenario in alleviating the mass budget problem previously mentioned.

The adoption of a larger $M_0$ (i.e. the model with $M_0=0.4$~\msun) at the same $M_\rmn{fin}$ reduces the total amount of polluted matter ($M_\rmn{acc}$) accreted on to the star. Combined with the more rapid formation of the radiative core, this leads to a lower helium abundance in the stellar core after the end of the late accretion phase, as clearly visible in Fig. \ref{fig:acc_dav_t0acc}.

Fig. \ref{fig:acc_dav_t0acc_hr} shows the quoted tracks in the \hr{} diagram, from the early \pms{} phase to an age of $10^8$~yr (approximatively the \zams). Close to the \zams, the models with late accretion are hotter and brighter than the standard ones with constant mass equal to $M_\rmn{fin}$, as a consequence of the higher helium abundance in the whole structure. However, the impact of changing the starting age $t_\rmn{0,l.acc}$ of the accretion is relevant mainly on the early \pms{} evolution.

Notice that if $t_\rmn{0,l.acc}=0$ is used, during the early \pms{} evolution, when models are still close to their Hayashi track, the tracks evolve during the accretion at an almost constant luminosity but at an increasing effective temperature. This feature is caused by the ignition of the deuterium burning during the accretion phase. If $t_\rmn{0,l.acc}\ge 2$~Myr is adopted, the d-burning ends before the beginning of the accretion and this particular feature is not visible in the \hr{} diagram (in this particular mass range).

It is worth noticing that all the presented accreting models (at a fixed $M_0$ and $M_\rmn{fin}$) attain approximatively the same luminosity and effective temperature close to the \zams{} (see Table \ref{tab:la_models}), even the models with the lowest central helium content (i.e. models with $t_\rmn{0,l.acc}=4$~Myr). In particular, the cases with $M_0=0.4$~\msun{} and $t_\rmn{0,l.acc}=4$~Myr have effective temperature slightly larger than those with the higher central helium abundance. This apparently strange behaviour is the consequence of the large amount of helium accreted in the envelope. Because of the fast formation of a radiative core, such models accrete the largest part of the polluted matter in the envelope and not in the centre. The large amount of helium in the envelope produces a local rise in the temperature (due to the increase of the molecular weight) which leads to an increase of the effective temperature.

Fig. \ref{fig:acc_dav_t0acc_li} shows the evolution of the surface $^7$Li abundance for the quoted late accretion models. The corresponding surface lithium content at the end of the late accretion and at the age of $10^8$ yr are listed in Table \ref{tab:la_models}. 

Standard tracks, i.e. models without any accretion episodes, predict only a mild \pms{} surface lithium depletion for stars of about 0.5 \msun{} while for 0.7 \msun{} no depletion occurs, because of the rapid development of a radiative core and of the consequent reduction of the temperature at the bottom of the convective envelope ($T_\rmn{ce}$). In accreting models, the evolution of the atmospheric lithium abundance is different because surface $^7$Li depletion depends on both the nuclear burning and the dilution with accreted lithium-poor material. The accretion process affects the structure of the star and in particular the temperature profile which in turn modifies both the lithium burning efficiency and the extension of the convective envelope. The latter influences the dilution degree of the survived original stellar lithium with lithium-free accreted material. The impact of changing the accretion starting age $t_\rmn{0,l.acc}$ on the surface $X_{^7\rmn{Li}}$ is thus twofold. First, $t_\rmn{0,l.acc}$ affects the onset and the efficiency of lithium-burning as it determines the beginning of the steep temperature rise induced by mass accretion (as the temperature strongly depends on the stellar mass). Secondly, $t_\rmn{0,l.acc}$ sets the start of the dilution process with lithium-depleted gas. At fixed $M_0$ and $M_\rmn{fin}$ values, if $t_\rmn{0,l.acc}$ increases the accretion begins on a fully convective star with a larger central temperature. Then, when the radiative core develops, the temperature at the base of the convective envelope ($T_\rmn{ce}$) -- at a given value of the total mass -- increases with $t_\rmn{0,l.acc}$. As the accretion proceeds, the convective envelope gets thinner and thinner and $T_\rmn{ce}$ decreases. However, the mass accretion leads to an increase of the  temperature along the whole structure. The two effects are opposite and partially counterbalance each other, as a consequence $T_\rmn{ce}$ remains almost constant during most of the accretion phase. The larger $T_\rmn{ce}$ in models with $t_\rmn{0,l.acc} > 0$ has the direct consequence of increasing the efficiency of $^7$Li-burning at the bottom of the convective envelope. Moreover, also the dilution process is more efficient in the models with larger $t_\rmn{0,l.acc}$ because of the lower mass contained in the convective envelope -- at a fixed total mass. Dilution and nuclear burning are more efficient in models with larger $t_\rmn{0,l.acc}$ causing a steeper surface $^7$Li depletion.

Regarding the dependence of $^7$Li surface content on the value of $M_0$, we note that if $M_0$ increases $T_\rmn{ce}$ decreases and the mass enclosed in the convective envelope increases. This reduces both the nuclear burning efficiency and the dilution leading to a larger $^7$Li surface content at the end of the accretion phase.

Fig. \ref{fig:acc_dav_t0acc_li} clearly shows that in the case of $M_0~=~0.2$~\msun{} and $t_\rmn{0,l.acc}~>~0$, independently of the final mass, lithium is completely destroyed within 6--8 Myr. On the other hand if $M_0~=~0.4$~\msun{} is used lithium is partially preserved. In the case $M_0=0.4$~\msun{} and $M_\rmn{fin}=0.5$~\msun, depending on the adopted $t_\rmn{0,l.acc}$, the resulting surface lithium depletion in the \zams{} can vary from about $-1.2$~dex ($t_\rmn{0,l.acc}$~=~0) to about $-2.7$~dex ($t_\rmn{0,l.acc}$~=~4~Myr). If $M_\rmn{fin}=0.5$~\msun{} is used then the depletion degree is smaller, namely $-0.5$ dex ($t_\rmn{0,l.acc}$~=~0) and $-1.6$ dex ($t_\rmn{0,l.acc}$~=~4~Myr). These values have to be compared to the results of standard non accreting models (at the same $M_\rmn{fin}$) that predict a mild (about $-1.3$ dex for $M~=~0.5$~\msun) or no $^7$Li depletion (for $M~=~0.7$~\msun). 

The analysis we performed confirms what already noticed by \citet{salaris14} and \citet{dantona14}, who showed that the accretion of a large amount of Li-free gas produces a lithium depletion much larger than observed. In agreement with what partially noted by \citet{salaris14}, the reduction of $t_\rmn{0,l.acc}$ could in part alleviate the discrepancy between the predicted and observed surface $^7$Li. Although not analysed here, we also recall that the increase of the duration of the accretion phase (i.e. $\Delta t_\mathrm{l.acc}$) leads to a more pronounced lithium depletion \citep{dantona14,salaris14}.

\section{Pre-MS models with both protostellar and late accretion}
\label{sec:psa}
A further step with respect to the \citet{dantona14} models consists in taking into account the early disc accretion phase during the protostellar evolution (\psa{} models) rather than starting the computation from the Hayashi track of the already formed stars. This means that the physical and chemical structure of a star with mass $M_0$ at the age $t_\rmn{0,l.acc}$ of the beginning of late accretion is computed consistently starting from the initial hydrostatic core of mass $M_\rmn{seed}$ ($\ll~M_0$) and following the protostellar mass accretion rather than from an arbitrary point on the Hayashi track of constant mass $M_0$. Nowadays there is a large consensus about the fact that low-mass stars gain a large fraction of their mass through accretion discs. 

This class of models is characterized by two distinct accretion phases, namely the protostellar one, which builds a star of mass $M_0$ starting from a seed of mass $M_\rmn{seed}$, and the late one, which leads to a star of final mass $M_\rmn{fin}$ accreting on an initial mass $M_0$. As already mentioned in Section \ref{sec:reference}, \psa{} models depend on some parameters (i.e. $\dot{m}$, $M_\rmn{seed}$). 

\citet{baraffe09} already showed that depending on the adopted $M_\rmn{seed}$ and $\dot{m}$, the \pms{} evolution of non-accreting and accreting models with the same final masses can be quite different, especially during the first 1--10~Myr. Hence, if the protostellar accretion phase affects the temporal duration of the fully convective phase, this might also have an effect on the amount of helium dragged down to the stellar centre, and consequently also on the subsequent evolution. 

However, such a first accretion phase relies on several poorly-constrained parameters. Thus, the aim of the following sections is to explore the effect of each of these parameters on the stellar characteristics during and at the end of the late accretion phase, and to analyse the differences with respect to the standard evolution and to the single late accretion \dv{} models described in Section \ref{sec:dv}.
 
\subsection{Effect of changing the seed mass}
\label{sec:mseed}
\begin{figure*}
	\centering
	\includegraphics[width=0.8\linewidth]{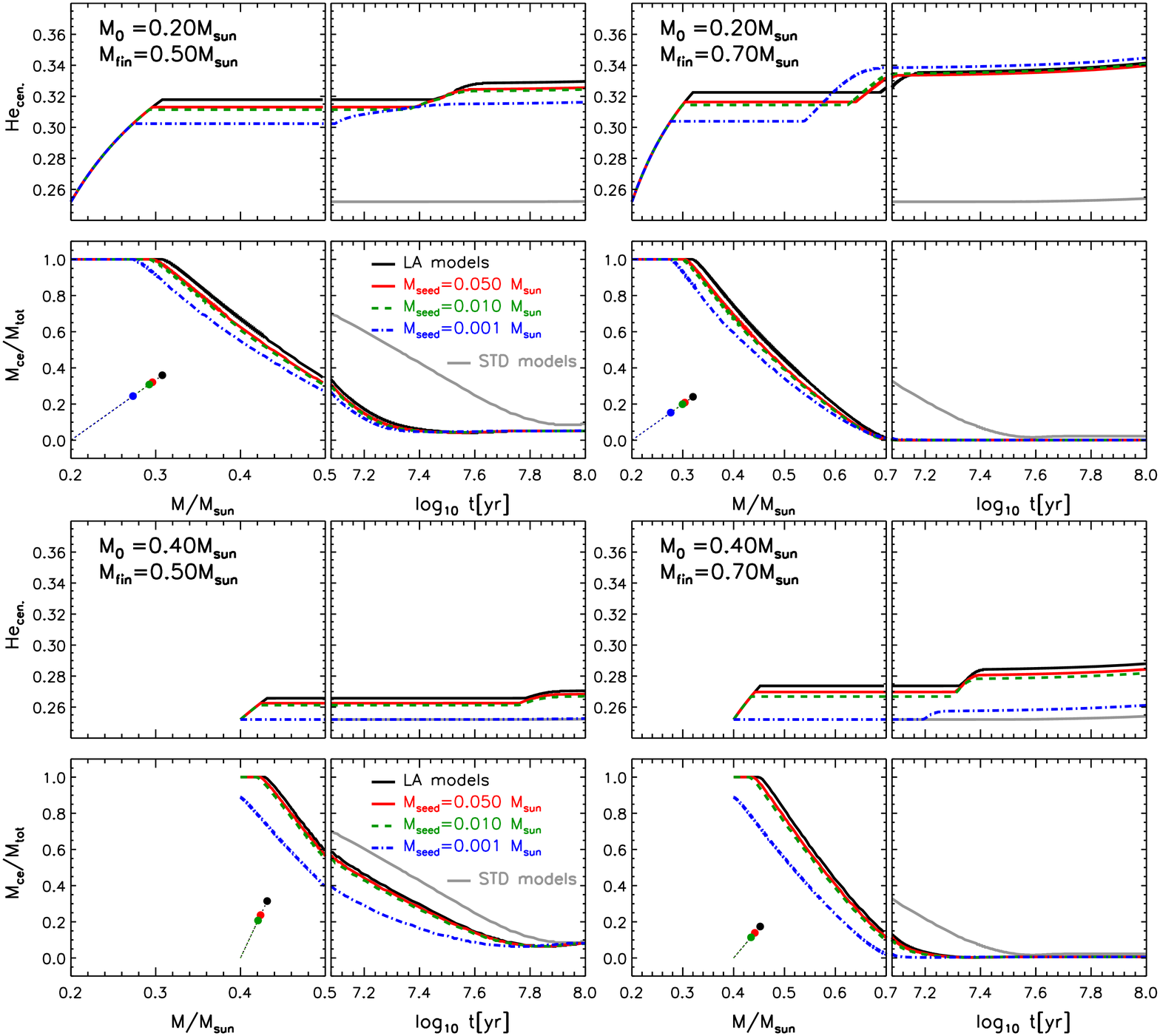}
	\caption{Evolution of the central helium mass fraction abundance (He$_\rmn{cen.}$) and of the fraction of stellar mass inside the convective envelope ($M_\rmn{ce}$/$M_\rmn{tot}$) as a function of both the total mass (during the accretion phase) and of the stellar age (for $t\ge12$~Myr) for the protostellar-late accretion models (\psa). Models are computed adopting the three labelled values of the seed mass ($M_\rmn{seed}$). Thin dotted lines in $M_\rmn{ce}$/$M_\rmn{tot}$ panel show the fraction of mass accreted during the fully convective phase (i.e. $M_\rmn{acc}^\rmn{fully conv.}/(M_\rmn{fin}-M_0)$); the end of such a phase is marked by a filled circle. Standard models (grey thick lines) with $M~=~M_\rmn{fin}$, $Y~=~Y_\rmn{ini}$ and $Z~=~Z_\rmn{ini}$ and \dv{} models (black thick lines) with $t_\rmn{0,l.acc}=2$~Myr are also shown. Top panels: models with initial mass $M_0~=~0.2$~\msun{} and two values of the final mass, namely $M_\rmn{fin}~=~0.5$ \msun{} (left-hand panels) and $M_\rmn{fin}~=~0.7$~\msun{} (right-hand panels).}
	\label{fig:acc_mini}
\end{figure*}
\begin{figure*}
	\centering
	\includegraphics[width=0.8\linewidth]{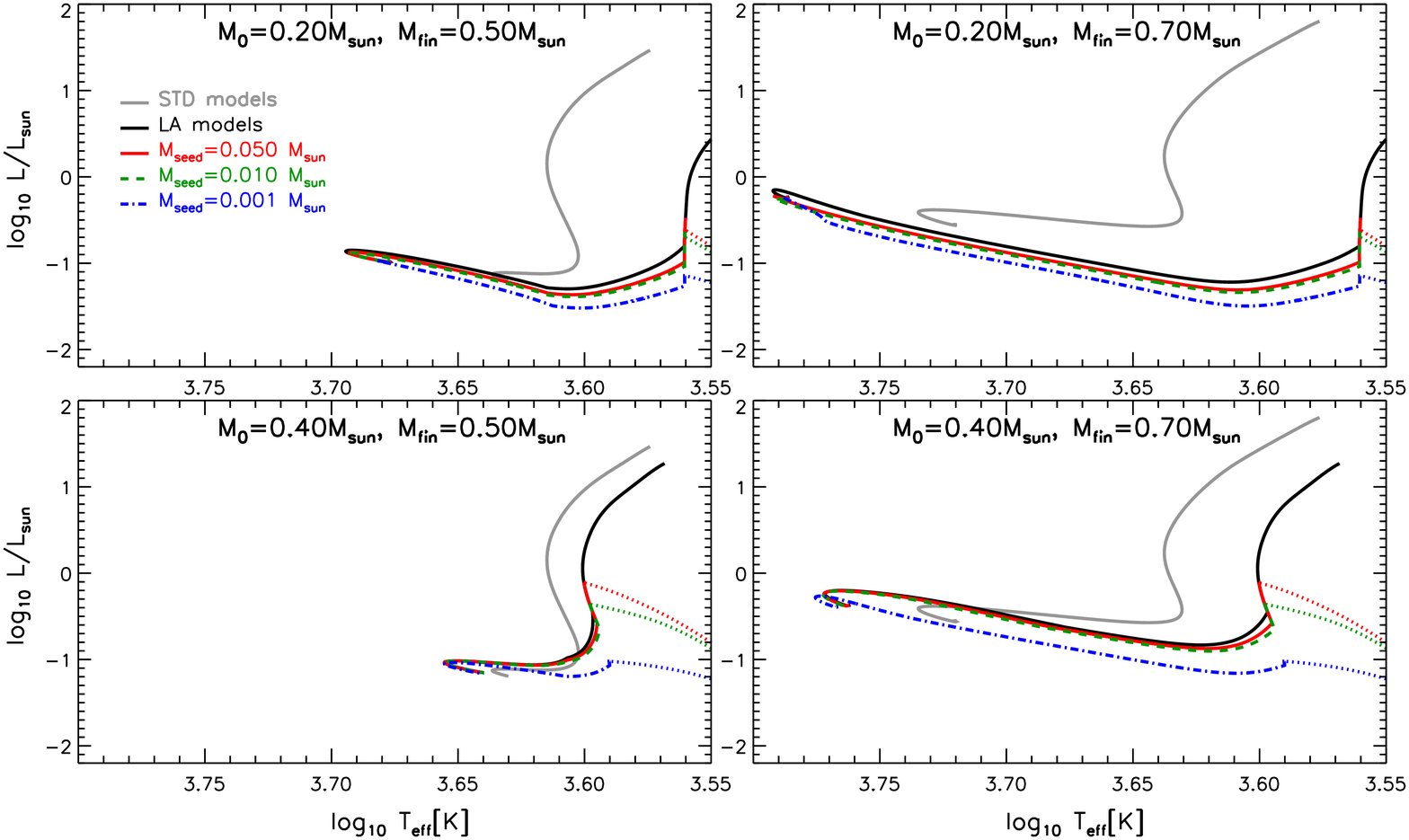}
	\caption{\hr{} diagrams for the protostellar-late accretion models (\psa) shown in Fig. \ref{fig:acc_mini}. Top panels: $M_0~=~0.2$~\msun{} with $M_\rmn{fin}~=~0.5$~\msun{} (left-hand panel) and $M_\rmn{fin}~=~0.7$~\msun{} (right-hand panel). The corresponding standard models are also shown (thick grey line). The dotted part of the evolutionary track represents the protostellar accretion phase. Bottom panel: the same as the top panel but for $M_0~=~0.4$~\msun.}
	\label{fig:acc_mini_hr}
\end{figure*}
\begin{figure*}
	\centering
	\includegraphics[width=0.8\linewidth]{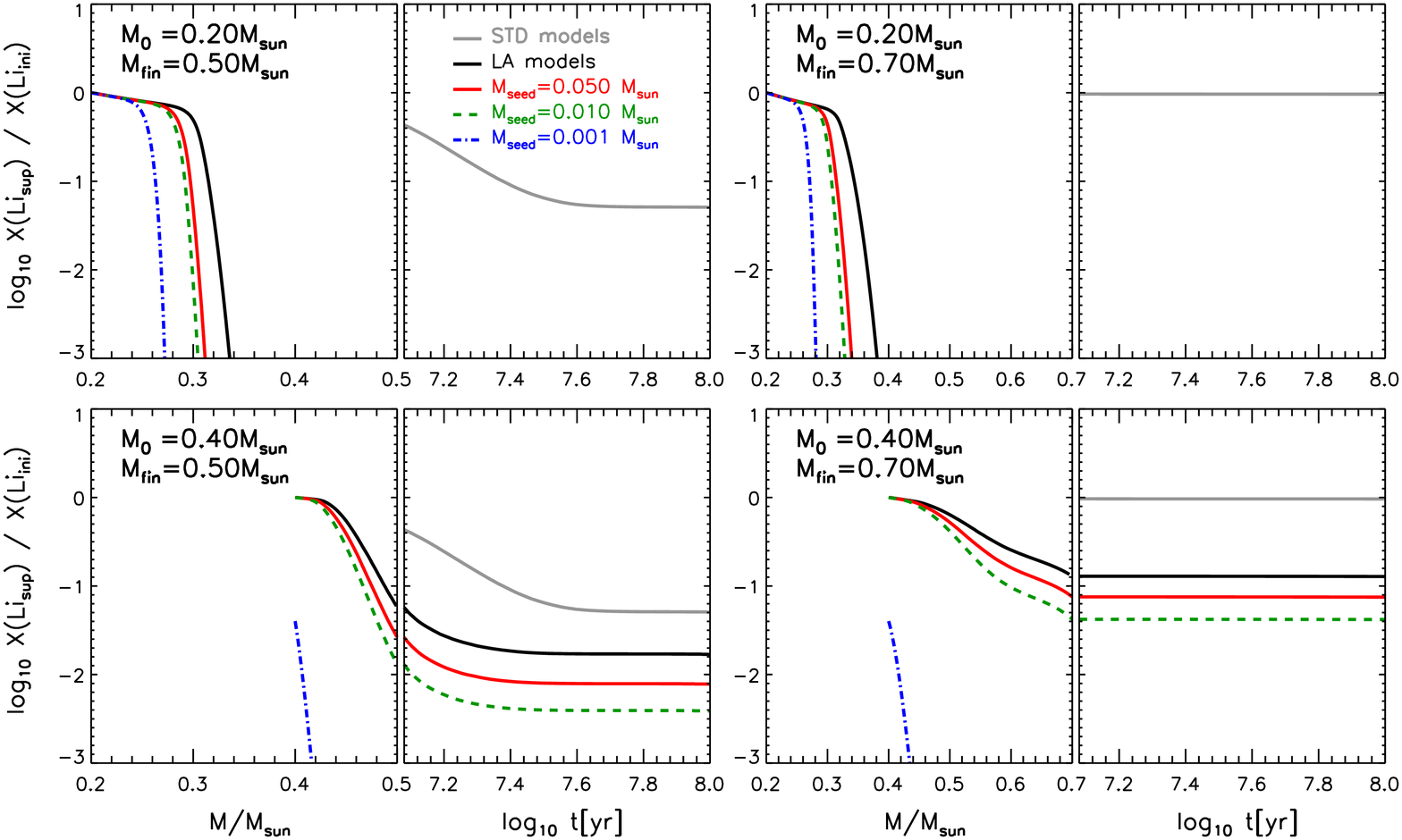}
	\caption{Surface $^7$Li abundance for the protostellar-late accretion models (\psa) shown in Fig. \ref{fig:acc_mini} as a function of both the total mass (during the accretion phase) and of the stellar age (for $t\ge12$~Myr). Top panels: $M_0~=~0.2$ \msun{} with $M_\rmn{fin}~=~0.5$~\msun{} (left-hand panel) and $M_\rmn{fin}~=~0.7$~\msun{} (right-hand panel). The corresponding standard models are also shown (thick grey line). Bottom panel: the same as the top panel but for $M_0~=~0.4$~\msun.}
	\label{fig:acc_mini_li}
\end{figure*}
Fig. \ref{fig:acc_mini} shows \psa{} models computed adopting three values of the seed mass, namely $M_\rmn{seed}~=~0.001$, 0.010, and 0.050~\msun, representative of the current mass range for the initial hydrostatic protostellar core \citep[see e.g.][]{masunaga00,boyd05,whitworth06,machida10,baraffe12}, and keeping fixed the other parameters to the reference values ($\dot{m}=10^{-5}$~\msun~yr$^{-1}$, $X_\rmn{d}=4\times 10^{-5}$, see Section \ref{sec:reference}). For comparison, the corresponding \dv{} tracks accounting only for the late accretion are also shown. Both the \dv{} and \psa{} models are computed for $t_{0,\rmn{l.acc}} = 2$~Myr. Notice that with the adopted $\dot{m}$ and $M_\rmn{seed}$ values the protostellar accretion phase ends at about 1.5--4$\times 10^{4}$~yr (depending on $M_0$), thus well before the late accretion beginning. The mass fractional abundance of central and surface helium at the end of the late accretion phase and at an age of $10^8$ yr for the quoted models are listed in Table \ref{tab:pla_models}.

Referring to Fig. \ref{fig:acc_mini}, when compared to the \dv{} models, the impact of protostellar accretion on the final results  is small for $M_\rmn{seed}~=~0.050$~\msun{} and gets larger and larger as the seed mass decreases. The lower $M_\rmn{seed}$ and the fainter and more compact, hence hotter, becomes the model left at the end of the protostellar accretion. For a given $M_0$ and $M_\rmn{fin}$, the radiative core develops at lower values of the stellar mass if $M_\rmn{seed}$ decreases and thus the final central helium abundance gets lower. Even in the most favourable case (i.e. $M_0~=~0.2$~\msun, top panels of Fig. \ref{fig:acc_mini}), the amount of helium accreted in the core in the \psa{} models is systematically lower ($Y~\approx~0.30$) than that of the corresponding \dv{} ($Y~\approx~0.32$). To be noted that the case $M_0~=~0.4$~\msun{} and $M_\rmn{seed}~=~0.001$~\msun{} (bottom panels of Fig. \ref{fig:acc_mini}) corresponds to the worst case, where the central helium abundance is not enhanced at all, because of the formation and the fast growth of the radiative core during the protostellar accretion phase. Such a radiative core continues to exist even after the beginning of the late accretion, thus preventing the accreted helium-rich material to reach the stellar centre. 

This clearly shows that if the lowest value of $M_\rmn{seed}~=~0.001$~\msun{} is adopted, there is a maximum value of $M_0$ ($0.2~\lt~M_0/~$\msun$~\la~0.4$) above which it is not possible to enhance the helium abundance in the stellar centre during the late accretion. In any case, the models which take into account protostellar evolution provide a central helium abundance at the end of the late accretion phase systematically lower than that of the models which neglected it.

For $M_0~=~0.2$~\msun, the development of a temporarily convective core allows the helium-rich matter accumulated in the outer envelope to be brought into the centre. The central helium abundance reached at $10^8$~yr is almost independent of $M_\rmn{seed}$ and is quite similar (although lower) to that in models without early accretion. Only for $M_0~=~0.2$~\msun{} and $M_\rmn{fin}~=~0.7$~\msun{} the model with the lowest $M_\rmn{seed}$, has a slightly larger central helium content (due to the formation of the temporarily convective core). The situation is different in the case of $M_0~=~0.4$~\msun; the adoption of the lowest value of $M_\rmn{seed}$ avoids the direct accretion of helium into the stellar core and even the temporarily formed convective core does not extend enough to bring a sizeable amount of helium rich material into the central regions.

Fig. \ref{fig:acc_mini_hr} shows the location in the \hr{} diagram of the tracks of the models which take into account protostellar accretion (\psa) together with those which follow only the late accretion (\dv) and standard \pms{} models of fixed mass equal to $M_\rmn{fin}$. As in the \dv{} cases discussed in Section \ref{sec:dv}, despite of the low central helium abundance, the $M_\rmn{seed}~=~0.001$~\msun{} model (lowest central helium abundance) is very close in \zams{} to the other accreting tracks, and even slightly hotter in the cases $M_0 = 0.4$ \msun{} and $M_\rmn{fin}=0.5$ and 0.7 \msun{} (see Table \ref{tab:pla_models}). As already mentioned such a behaviour is caused by the large amount of helium deposed in the external convective envelope, after the formation of the radiative core.

The change of $M_\rmn{seed}$ affects the \pms{} evolution of \psa{} models, as clearly visible in Fig. \ref{fig:acc_mini_hr}. Increasing the seed mass, the point at the end of the protostellar accretion phase (the end of the dotted line) is located progressively at higher luminosity and closer to the Hayashi track of the \dv{} model which does not take into account protostellar accretion with the same $M_0$, $M_\rmn{fin}$, and $t_\rmn{0,l.acc}$. Nevertheless, in the \dv{} set of models the late accretion episode begins at luminosities systematically higher than in the \psa{} set of models, even though in both cases the late accretion occurs at the same age, i.e. $t_\rmn{0,acc}~=~2$~Myr. The different luminosity (or more in general the different location in the \hr{} digram) at the end of the first accretion phase highlights the fact that including or not the protostellar accretion leads to \pms{} models with different structures at a similar age \citep[see e.g.][]{baraffe09,hosokawa11}.

Fig. \ref{fig:acc_mini_li} shows the surface $^7$Li abundance evolution for the \psa{} models computed with the labelled seed masses. The corresponding surface lithium content at the end of the late accretion phase and at the age of $10^8$ yr are listed in Table \ref{tab:pla_models}. It is clearly visible that changing $M_\rmn{seed}$ has a drastic effect on the predicted surface lithium abundance. During the protostellar accretion phase the atmospheric lithium content does not appreciably change due to both the low nuclear burning efficiency at the typical temperatures characterizing this phase and the concomitant continuous supply of fresh $^7$Li in accreted matter. Such a supply stops at the end of the protostellar accretion and does not start again during the late accretion as the helium-rich polluting gas is fully depleted of lithium. Thus, the increasing efficiency of Li-burning with time and, then, the dilution with Li-free gas during the late accretion lead to a progressive depletion of atmospheric lithium. 

The change of $M_\rmn{seed}$ affects both the temperature at the base of the convective envelope and the mass contained in it, both at the end of the protostellar accretion and during the late accretion phase. The lower is $M_\rmn{seed}$ and the larger is $T_\rmn{ce}$ and the lower is $M_\rmn{ce}$ at a given total mass. Similarly to what discussed in the previous section, this leads to both a more efficient $^7$Li nuclear burning and dilution hence to a lower surface lithium content at a given mass. The adoption of $M_\rmn{seed}~=~0.001$ \msun{} leads to the complete destruction of surface $^7$Li within 3--6 Myr in all the analysed models. In the case of $M_0~=~0.2$~\msun, similarly to the \dv{} case, lithium is completely destroyed within 6~Myr even if larger $M_\rmn{seed}$ are adopted. In the case of $M_0~=~0.4$~\msun, the nuclear burning and dilution are less efficient and surface $^7$Li is partially preserved, but only if $M_\rmn{seed} > 0.001$~\msun. If M$_\rmn{seed}~=~0.001$~\msun{} and M$_0~=~0.4$~\msun{} is adopted, $^7$Li is destroyed quite efficiently even at the end of the protostellar accretion, and the star is about $-1.4$ dex $^7$Li depleted at the beginning of late accretion, when the residual $^7$Li is rapidly destroyed. In all the presented cases the resulting $^7$Li abundance is systematically lower than that of the  corresponding models which do not take into account protostellar accretion (\dv) or standard models (\std).

In summary, if a realistic value $M_\rmn{seed}~\approx~0.001$~\msun{} is adopted for the initial hydrostatic core, the inclusion of the protostellar evolution in stellar model computation significantly affects also the theoretical predictions of more advanced evolutionary phases, in particular the amount of helium rich material carried into the stellar core and the surface $^7$Li abundance.

\subsection{Effect of changing the mass accretion rate}
\label{sec:mdot}
\begin{figure*}
	\centering
	\includegraphics[width=0.8\linewidth]{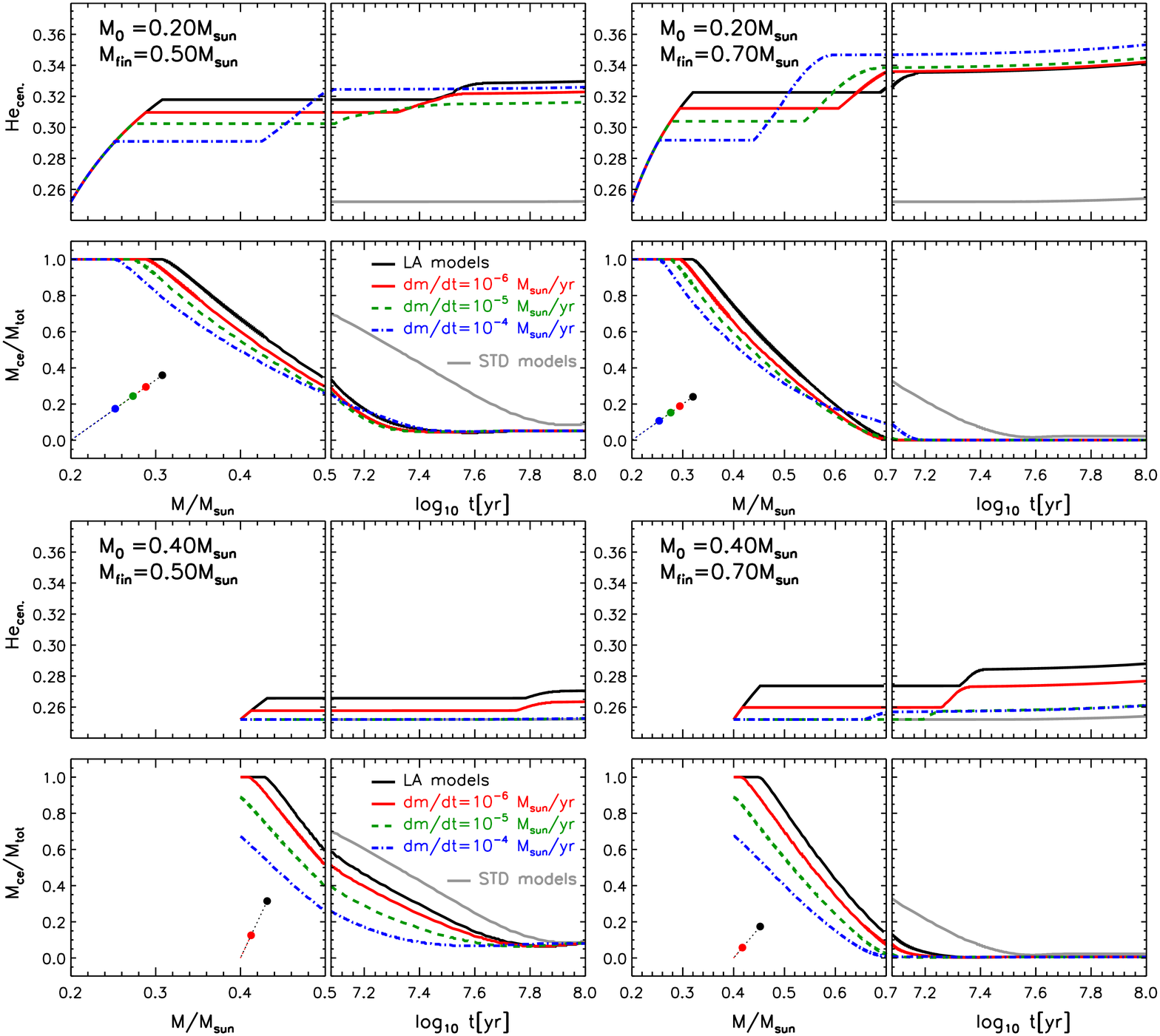}
	\caption{As in Fig. \ref{fig:acc_mini} but for models computed adopting the three labelled protostellar mass accretion rate values ($\dot{m}$).}
	\label{fig:acc_mdot}
\end{figure*}
\begin{figure*}
	\centering
	\includegraphics[width=0.8\linewidth]{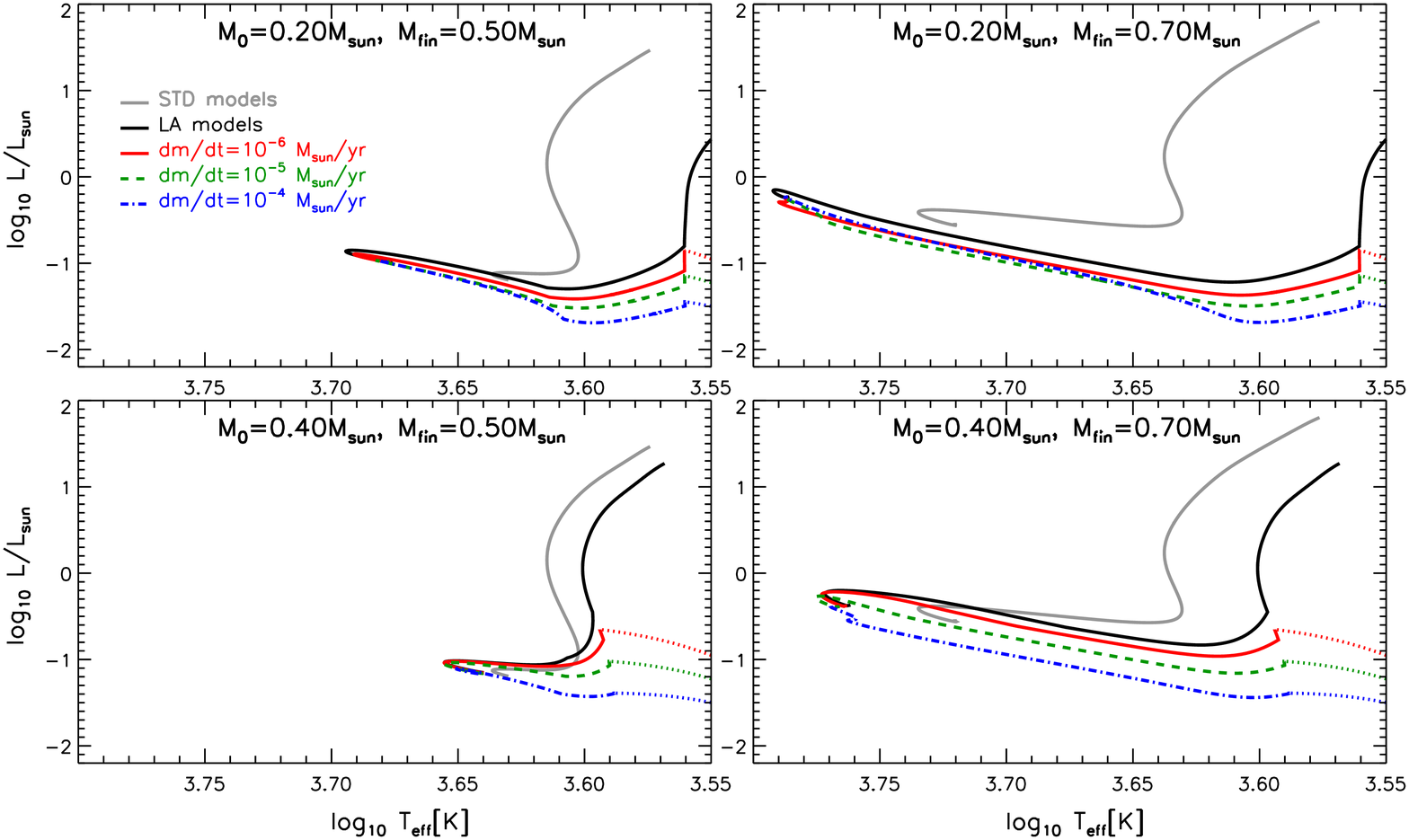}
	\caption{\hr{} diagrams for the protostellar-late accretion models (\psa) shown in Fig. \ref{fig:acc_mdot}. Top panels: $M_0~=~0.2$~\msun{} with $M_\rmn{fin}~=~0.5$~\msun{} (left-hand panel) and $M_\rmn{fin}~=~0.7$~\msun{} (right-hand panel). The corresponding standard models are also shown (thick grey line). The  dotted part of the evolutionary track represents the protostellar accretion phase. Bottom panel: the same as the top panel but for $M_0~=~0.4$~\msun.}
	\label{fig:acc_mdot_hr}
\end{figure*}
\begin{figure*}
	\centering
	\includegraphics[width=0.8\linewidth]{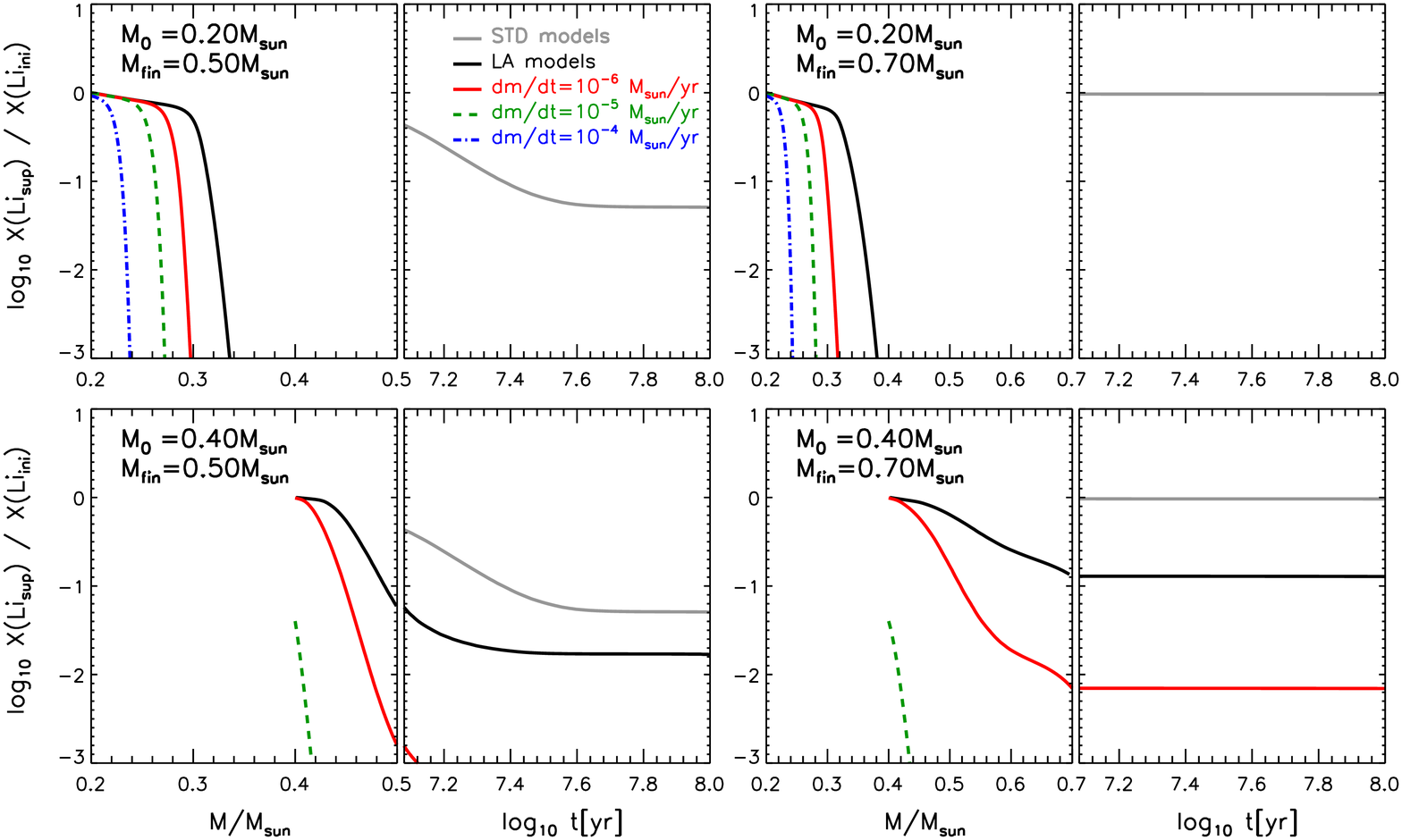}
	\caption{Surface $^7$Li abundance for the protostellar-late accretion models (\psa) shown in Fig. \ref{fig:acc_mdot} as a function of both the total mass (during the accretion phase) and of the stellar age (for $t\ge12$~Myr). Top panels: $M_0~=~0.2$~\msun{} with $M_\rmn{fin}~=~0.5$~\msun{} (left-hand panel) and $M_\rmn{fin}~=~0.7$~\msun{} (right-hand panel). The corresponding standard models are also shown (thick grey line). Bottom panel: the same as the top panel but for $M_0~=~0.4$~\msun.}
	\label{fig:acc_mdot_li}
\end{figure*}

Fig. \ref{fig:acc_mdot} shows the evolution of the central helium content and of the convective envelope extension of the \psa{} models computed adopting three protostellar accretion rate values, namely $\dot{m}~=~10^{-6},\,10^{-5},$ and $10^{-4}$~\msun~$\rmn{yr}^{-1}$ \citep[see e.g][ and references therein]{baraffe09,baraffe12}, compared with the standard and \dv{} models for the labelled $M_0$ and $M_\rmn{fin}$ values. For the computation of the \psa{} models the reference seed mass ($M_\rmn{seed}=0.001$~\msun), initial deuterium abundance ($X_\rmn{d}=4\times10^{-5}$) and $t_{0,\rmn{l.acc.}} = 2$~Myr have been used. The mass fractional abundance of central and surface helium at the end of the late accretion phase and at an age of $10^8$ yr for the quoted models are listed in Table \ref{tab:pla_models}.

The mass accretion rate has a relevant effect on the structure of accreting objects and on the formation of the radiative core. The higher is the accretion rate and the faster is the contraction, thus the heating of the star. Depending on $M_0$ and $\dot{m}$, the star develops, in the worst case, a radiative core before the end of the protostellar accretion phase which is still present during the late accretion (i.e. $M_0~=~0.4$~\msun{} and $\dot{m}~\ge~10^{-5}$~\msun~$\rmn{yr}^{-1}$). For $M_0~=~0.2$~\msun, the star remains fully convective during the whole protostellar accretion phase, independently of $\dot{m}$. However, the duration of the fully convective phase during the late accretion episode progressively reduces as $\dot{m}$ increases. As a result, at the end of the late accretion, the central helium content of the protostellar accretion models, is lower than that of the corresponding \dv{} ones which neglected it. 

The increase of the central helium abundance caused by the development of a convective core depends on the structure of the star left at the end of the protostellar accretion, thus on the mass accretion rate. In the most favourable case, which  corresponds to $M_0~=~0.2$~\msun{} and $M_\rmn{fin}~=~0.7$~\msun, all the quoted accretion rates lead to a central helium abundance higher than that obtained in the \dv{} models. On the other hand, if $M_0~=~0.4$~\msun{} is used then the larger is $\dot{m}$ and the lower is the central helium. In this case the central helium abundance is systematically lower than that of the corresponding \dv{} models.

Fig. \ref{fig:acc_mdot_hr} shows the \hr{} diagram of the quoted models. As expected, the impact on the evolutionary track of the accretion rate adopted during the protostellar phase is particularly evident in \pms{} phase; the higher is $\dot{m}$ and the more compact and fainter is the star at the end of the early accretion phase. Even in this case, all the tracks have a similar location in the \hr{} diagram close to the \zams, independently of the central helium content. The model with the lower helium abundance in the core, but the higher in the external envelope (i.e. the model with $\dot{m}=10^{-4}$~\msun~yr$^{-1}$), is the hottest in the \hr{} diagram close to the \zams{} (see Table \ref{tab:pla_models}). 

Fig. \ref{fig:acc_mdot_li} shows the surface $^7$Li abundance evolution. The corresponding surface lithium content at the end of the late accretion phase and at the age of $10^8$ yr are listed in Table \ref{tab:pla_models}. As shown in Section \ref{sec:mseed}, the models which take into account protostellar accretion predict a surface lithium abundance lower than that of the models which neglect it. Moreover, the higher is $\dot{m}$ and the larger the lithium depletion at a given total mass and at the end of the late accretion. Indeed, an increase of $\dot{m}$ results in a large temperature at the bottom of the convective envelope and in a lower $M_\rmn{ce}$ values, thus increasing both the $^7$Li nuclear burning and dilution efficiency. In the case of $M_0~=~0.2$ \msun, the \psa{} models completely deplete their lithium within few Myr, at progressively younger ages as $\dot{m}$ increases. In both the cases with $M_0~=~0.4$~\msun{} ($M_\rmn{fin} = 0.5$ and $0.7$ \msun) lithium is partially preserved if a low accretion rate is adopted ($\dot{m}~=~10^{-6}$~\msun~$\rmn{yr}^{-1}$), while it is completely destroyed if larger $\dot{m}$ values are used. Notice that, the worst case corresponds to $M_0~=~0.4$ \msun{} and $\dot{m}~=~10^{-4}$~\msun~$\rmn{yr}^{-1}$, where $^7$Li is completely and efficiently destroyed before the beginning of the late accretion phase. 

\subsection{Effect of changing the initial deuterium abundance}
\label{sec:xd}
\begin{figure*}
	\centering
	\includegraphics[width=0.8\linewidth]{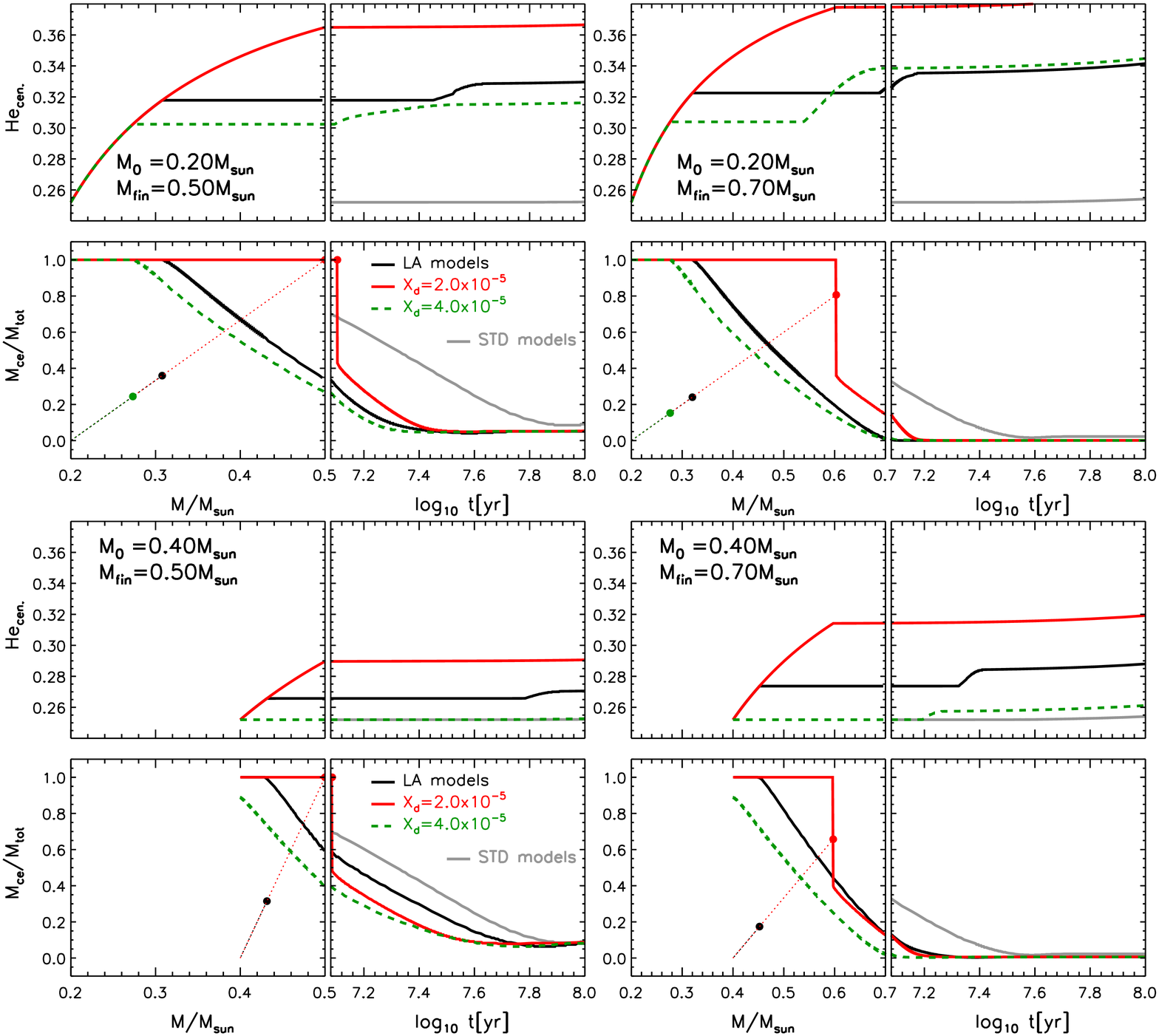}
	\caption{As in Fig. \ref{fig:acc_mini} but for models computed adopting the two labelled initial deuterium abundance values ($X_\rmn{d}$).}
	\label{fig:acc_xd}
\end{figure*}
\begin{figure*}
	\centering
	\includegraphics[width=0.8\linewidth]{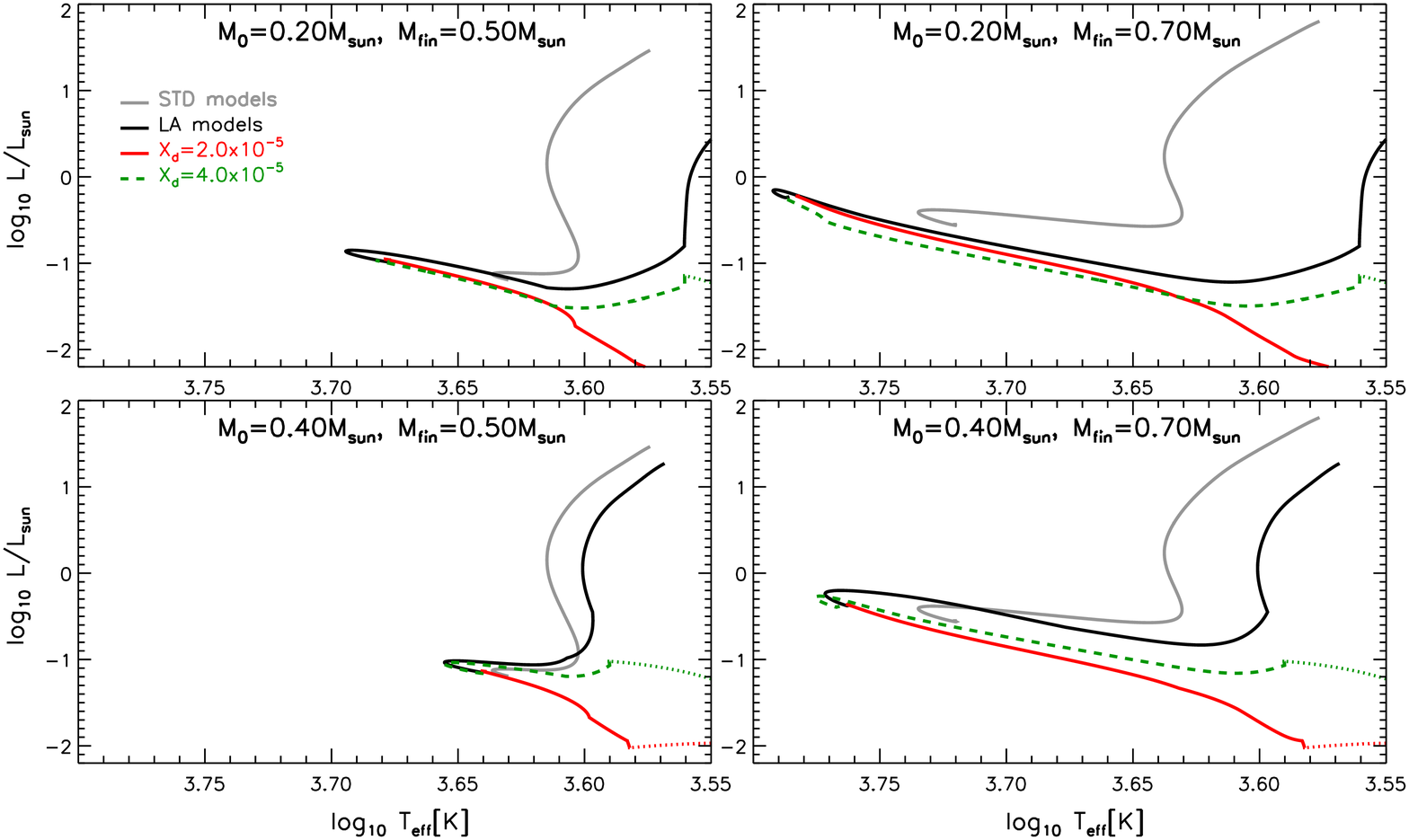}
	\caption{\hr{} diagrams for the protostellar-late accretion models (\psa) shown in Fig. \ref{fig:acc_xd}. Top panels: $M_0~=~0.2$~\msun{} with $M_\rmn{fin}~=~0.5$~\msun{} (left-hand panel) and $M_\rmn{fin}~=~0.7$~\msun{} (right-hand panel). The corresponding standard models are also shown (thick grey line). The  dotted part of the evolutionary track represents the protostellar accretion phase. Bottom panel: the same as the top panel but for $M_0~=~0.4$~\msun.}
	\label{fig:acc_xd_hr}
\end{figure*}
\begin{figure*}
	\centering
	\includegraphics[width=0.8\linewidth]{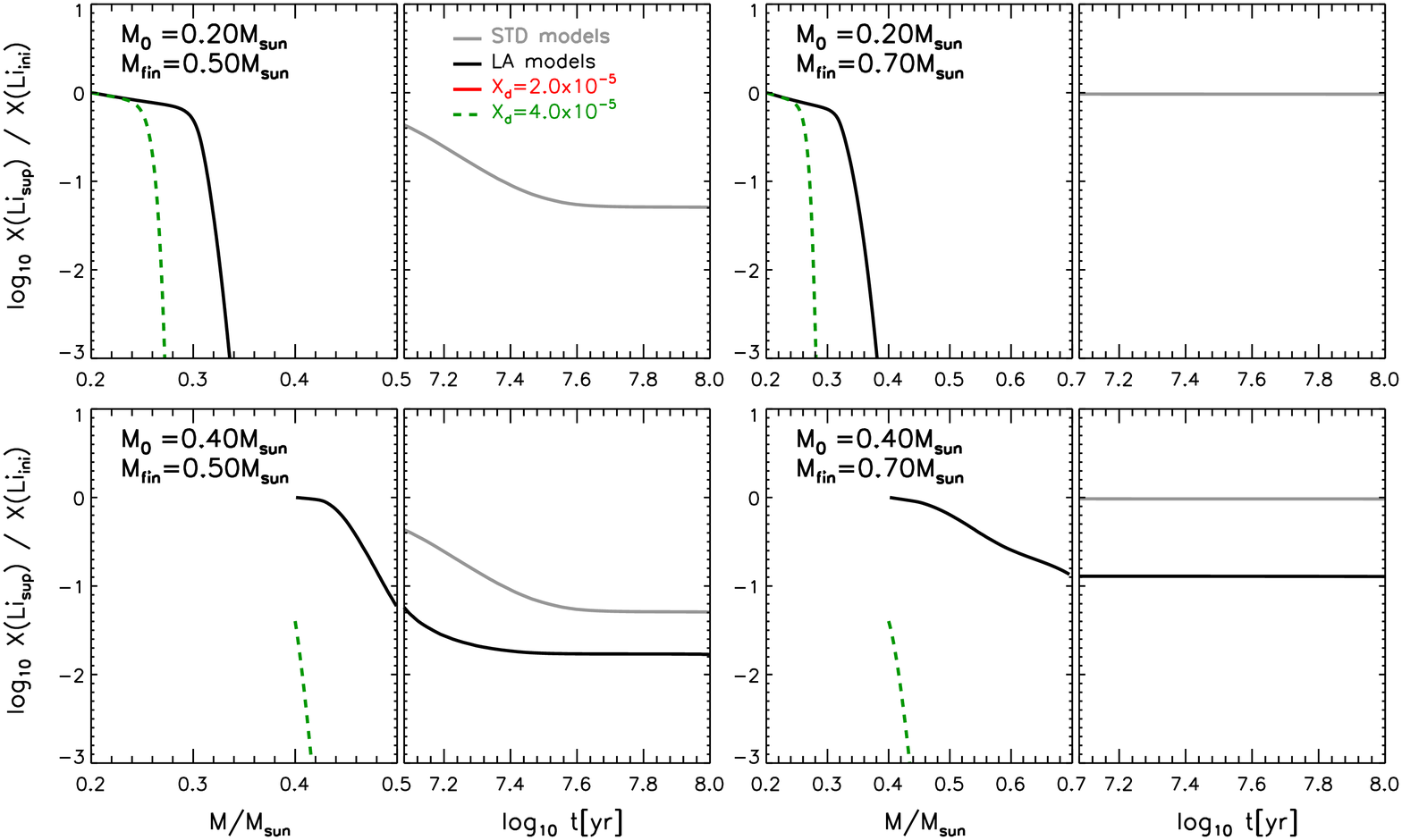}
	\caption{Surface $^7$Li abundance for the protostellar-late accretion models (\psa) shown in Fig. \ref{fig:acc_xd} as a function of both the total mass (during the accretion phase) and of the stellar age (for $t\ge12$~Myr). Top panels: $M_0~=~0.2$~\msun{} with $M_\rmn{fin}~=~0.5$~\msun{} (left-hand panel) and $M_\rmn{fin}~=~0.7$~\msun{} (right-hand panel). The corresponding standard models are also shown (thick grey line). Bottom panel: the same as the top panel but for $M_0~=~0.4$~\msun.}
	\label{fig:acc_xd_li}
\end{figure*}

Deuterium burning plays a crucial role during the standard \pms{} evolution, as it ignites at relatively low temperature (about 1 million degree) and it generates energy enough to temporarily balance the radiative losses at the stellar surface; this produces a significant slowdown in the rate of gravitational contraction. In accreting models, the continuous mass growth requires progressively more efficient d-burning to keep the stellar structure stable. However, depending on the deuterium abundance of the disc gas, this condition might not be fulfilled during the whole accretion phase, with the consequence that the accreting star starts to contract again.

The initial deuterium abundance in the molecular cloud from which the star forms can range from about $X_\rmn{d}~\approx~4 \times 10^{-5}$ \citep[primordial value,][]{pettini08,cooke14} to $X_\rmn{d}~\approx~2\times 10^{-5}$ \citep[solar neighbourhood;][]{sembach10}. In order to cover this abundance range, we adopted $X_\rmn{d}~=~2\times 10^{-5}$ and $X_\rmn{d}~=~4\times 10^{-5}$ (our reference). We used a deuterium abundance different from zero only during the protostellar accretion phase, while during the late accretion the material is supposed to be fully deuterium-depleted.

Fig. \ref{fig:acc_xd} shows the effect of the adopted deuterium abundance on the central helium content and the convective envelope mass fraction during the late accretion phase. The mass fractional abundance of central and surface helium at the end of the late accretion phase and at an age of $10^8$ yr for the quoted models are listed in Table \ref{tab:pla_models}. A reduction of the initial deuterium abundance from $X_\rmn{d}=4\times10^{-5}$ to $X_\rmn{d}=2\times10^{-5}$ during the protostellar accretion phase leads to a more compact structure with a larger central temperature. In the $X_\rmn{d}=2\times10^{-5}$ models a central temperature of about $T \sim 7\times 10^6$K typical of the ignition of the hydrogen burning can be achieved even at the end of the protostellar accretion. Such a large temperature continues to increase during the late accretion, reaching about $10^7$~K. This value is then kept constant during the largest part of the late accretion. The large energy flux generated by the hydrogen burning reactions at the centre of the star induces central convection. Moreover the external regions are cold enough to have a large opacity, and the star continues to be fully convective for all the protostellar and part of the late accretion. An external convective envelope only forms when the total mass reaches 0.6~\msun. Given such a situation, independently of $M_0$, $X_\rmn{d}~=~2\times 10^{-5}$ models show a larger central helium content than both \psa{} models with $X_\rmn{d}~=~4\times 10^{-5}$ and \dv{} ones which do not take into account protostellar accretion. We recall that the \dv{} models have been computed using $X_\rmn{d}~=~4\times 10^{-5}$.

Fig. \ref{fig:acc_xd_hr} shows the \hr{} diagram of the accreting models computed adopting the two quoted values for the initial deuterium abundance. It is clearly visible the large impact on the \pms{} evolution. Moreover, the variation of $X_\rmn{d}$ has a significant effect in determining also the position of models close to the \zams{} (especially in the case of $M_0~=~0.4$~\msun). The track with the largest helium content in the envelope (i.e. $X_\rmn{d}~=~4\times10^{-5}$) results to be hotter in \zams{} than the others (see Table \ref{tab:pla_models}), although the central helium abundance is lower.

Fig. \ref{fig:acc_xd_li} shows the surface $^7$Li abundance evolution. The corresponding surface lithium content at the end of the late accretion phase and at the age of $10^8$ yr are listed in Table \ref{tab:pla_models}. The surface abundances corresponding to the $X_\rmn{d}~=~2\times 10^{-5}$ are not visible in figure because lithium is completely destroyed during the protostellar accretion phase. This is the consequence of the deuterium-poor models reaching central temperatures much larger ($\sim 10^7$K, about one order of magnitude) than those of deuterium rich stars during the protostellar accretion phase in fully convective structures. This eventually leads to increase enough the efficiency of lithium-burning to fully deplete surface $^7$Li before protostellar accretion ends. Consequently, no lithium survives if $X_\rmn{d}=2\times10^{-5}$ is adopted in \psa{} models.

\subsection{Effect of changing the age of the beginning of the late accretion}
\label{sec:t0acc}
\begin{figure*}
	\centering
	\includegraphics[width=0.8\linewidth]{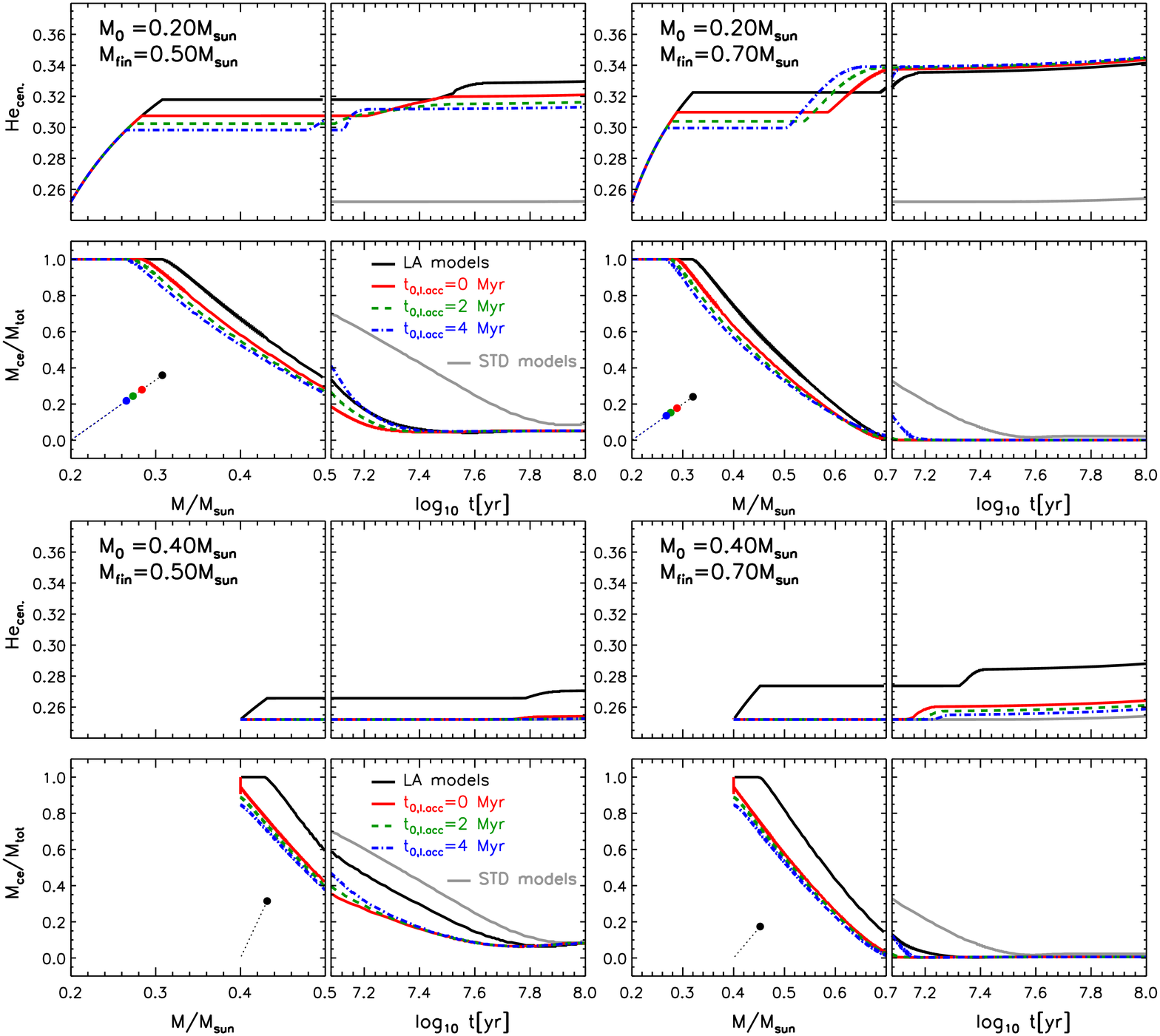}
	\caption{As in Fig. \ref{fig:acc_mini} but for models computed adopting the three labelled values of the beginning of the late accretion  starting ages ($t_\rmn{0,l.acc}$).}
	\label{fig:acc_t0acc}
\end{figure*}
\begin{figure*}
	\centering
	\includegraphics[width=0.8\linewidth]{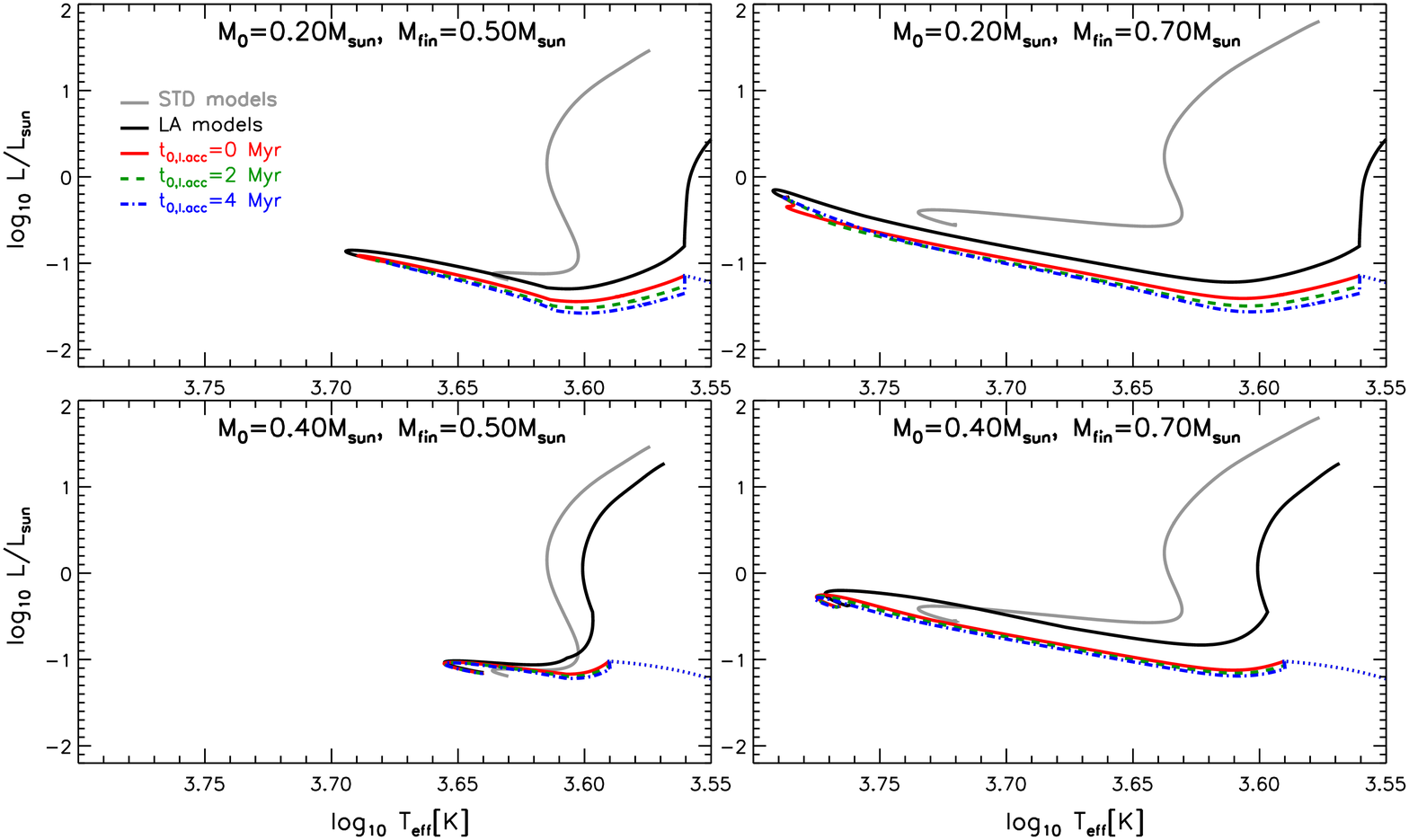}
	\caption{\hr{} diagrams for the protostellar-late accretion models (\psa) shown in Fig. \ref{fig:acc_t0acc}. Top panels: $M_0 = 0.2$ \msun{} with $M_\rmn{fin} = 0.5$ \msun{} (left-hand panel) and $M_\rmn{fin} = 0.7$ \msun{} (right-hand panel). The corresponding standard models are also shown (thick grey line). The   dotted part of the evolutionary track represents the protostellar accretion phase. Bottom panel: the same as the top panel but for $M_0 = 0.4$ \msun.}
	\label{fig:acc_t0acc_hr}
\end{figure*}
\begin{figure*}
	\centering
	\includegraphics[width=0.8\linewidth]{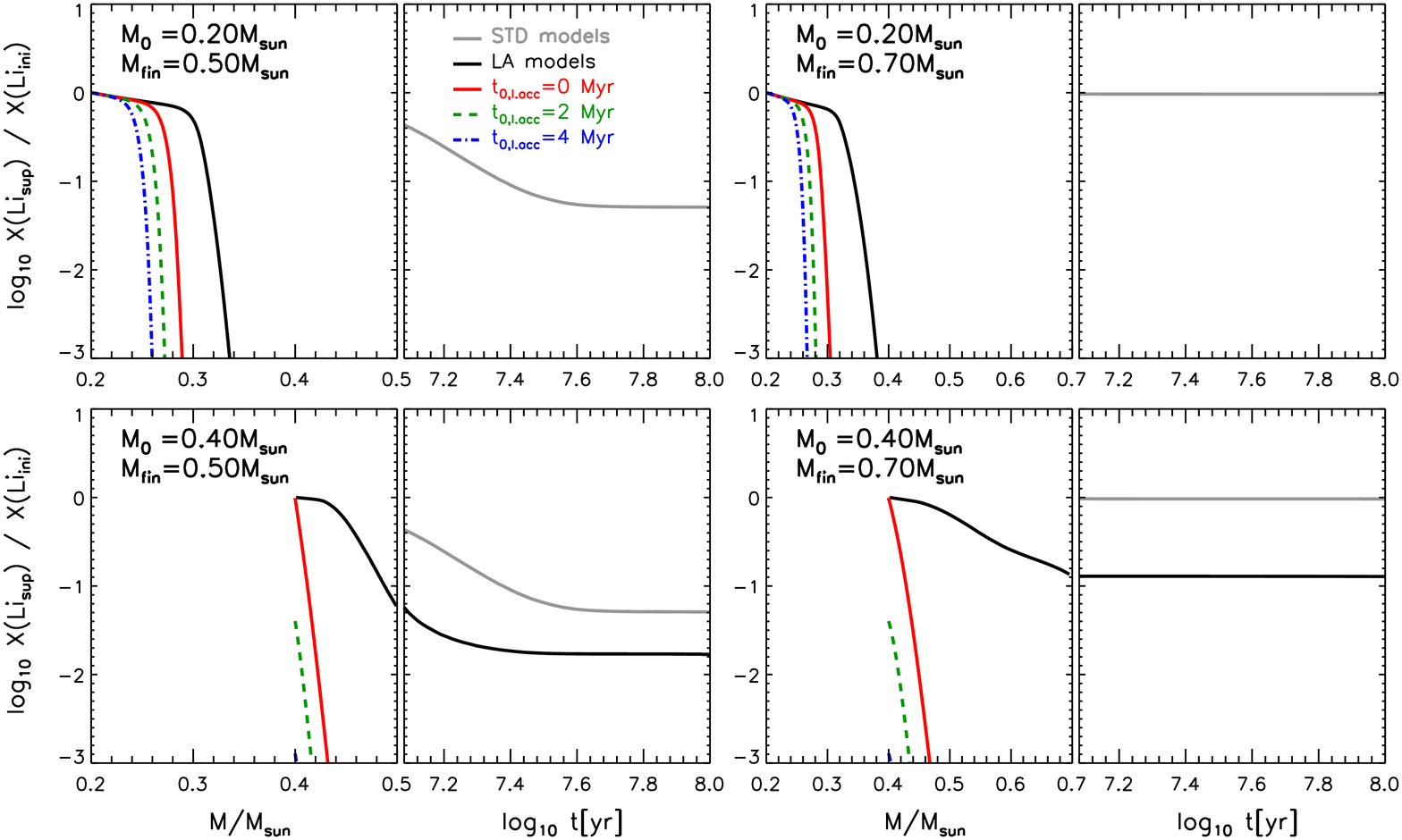}
	\caption{Surface $^7$Li abundance for the protostellar-late accretion models (\psa) shown in Fig. \ref{fig:acc_t0acc} as a function of both the total mass (during the accretion phase) and of the stellar age (for $t\ge12$~Myr). Top panels: $M_0~=~0.2$~\msun{} with $M_\rmn{fin}~=~0.5$~\msun{} (left-hand panel) and $M_\rmn{fin}~=~0.7$~\msun{} (right-hand panel). The corresponding standard models are also shown (thick grey line). Bottom panel: the same as the top panel but for $M_0~=~0.4$~\msun.}
	\label{fig:acc_t0acc_li}
\end{figure*}

In Section \ref{sec:dv}, we discussed the effect of changing the starting age $t_\rmn{0,l.acc}$ of the late accretion episode in \dv{} models. In this section, we performed a similar comparison but for the \psa{} models, which take into account protostellar accretion. We adopted the reference $M_\rmn{seed} = 0.001$~\msun, $\dot{m} = 10^{-5}$~\msun~yr$^{-1}$ and $X_\rmn{d}=4\times10^{-5}$ but three different time for the beginning of the late accretion, namely $t_\rmn{0,l.acc}~=~0, \, 2,$ and 4~Myr. The case $t_\rmn{0,l.acc}~=~0$ does not mean that the late accretion begins at the stellar age $t=0$; indeed, at  $t=0$ the star is experiencing the protostellar accretion phase. Thus, in this case $t_\rmn{0,l.acc}~=~0$ means that the late accretion starts right after the end of the first accretion phase at ages much younger than 1 Myr ($\la 4\times10^4$ yr, depending on $M_0$). 

As one can see in Fig. \ref{fig:acc_t0acc}, the fully convective phase gets slightly shorter if $t_\rmn{0,l.acc}$ increases, thus reducing the amount of polluted matter actually channelled into the centre and consequently the final central helium abundance. However, at variance with the results discussed in Section \ref{sec:dv}, models that take into account protostellar accretion are almost unaffected by a change of $t_\rmn{0,l.acc}$. This occurs because the star left at the end of the protostellar accretion phase is relative compact and faint, with typical Kelvin--Helmholtz time-scale ($\tau_\rmn{KH}$) of about 10 Myr. Since in these cases $\tau_\rmn{KH}$ is larger than the variation of $t_\rmn{0,l.acc}$ (i.e. 0--4 Myr) the stellar structure on which accretion occurs is rather insensitive to a variation of $t_\rmn{0,l.acc}$. As a consequence the late accretion is only marginally affected by the adopted $t_\rmn{0,l.acc}$.

As in the previous \psa{} cases, only for $M_0~=~0.2$~\msun{} the star remains fully convective for, at least, part of the late accretion. In the other case ($M_0~=~0.4$~\msun) the development of the radiative core prevents helium rich matter to reach the stellar centre. The formation of the convective core allows the previously helium enhanced matter accumulated in the envelope to be brought into the centre producing a growth of central helium abundance. The adoption of $t_\rmn{0,l.acc}=0$ results in a slightly larger central helium content in all the cases, but systematically lower than that obtained in the case of the \dv{} models. However, the impact of $t_\rmn{0,l.acc}$ on the \psa{} models is only marginal for what concerns both the central helium abundance at the end of the late accretion phase and the morphology of the tracks in the \hr{} diagram, as shown in Fig. \ref{fig:acc_t0acc_hr}.

Fig. \ref{fig:acc_t0acc_li} shows the surface $^7$Li abundance evolution for \psa{} models computed with the labelled $t_\rmn{0,l.acc}$. The corresponding surface lithium content at the end of the late accretion phase and at the age of $10^8$ yr are listed in Table \ref{tab:pla_models}. As already discussed in the Sections \ref{sec:mseed}--\ref{sec:xd}, the adoption of a small initial seed mass, i.e. $M_\rmn{seed}~=~0.001$~\msun, leads to a very rapid destruction of lithium, independently of $M_0$ and/or $M_\rmn{fin}$. The age at which lithium is fully depleted only marginally depends on $t_\rmn{0,l.acc}$ too. Indeed, the use of different $t_\rmn{0,l.acc}$ only slightly modifies the temperature at the bottom of convective envelope (which slightly increases with $t_\rmn{0,l.acc}$) and the extension of the external convective region (which decreased with $t_\rmn{0,l.acc}$). This results in a lower surface lithium content during the late accretion at a given mass if $t_\rmn{0,l.acc}$ increases. In any case surface lithium is completely destroyed within 5--6 Myr, and in the case of $M~=~0.4$~\msun{} and $t_\rmn{0,l.acc}~=~4$~Myr $^7$Li is completely destroyed before the beginning of the late accretion phase.

\section{Conclusions}
\label{sec:conclusions}
\citet{bastian13} have recently suggested that accretion of helium-rich gas ejected by interactive massive binary systems on low and very-low mass \pms{} stars could be a viable scenario to form multiple populations in \gcs. Following the first analysis by \citet{cassisi14}, \citet{salaris14}, and \citet{dantona14}, we computed \pms{} models which take into account  the accretion of polluted matter in order to quantify the effect on the evolution of central helium and surface lithium abundances. The main novelties with respect to the quoted studies is the analysis of the effect of: (i) varying the starting age of the late accretion of polluted matter during the \pms{} phase,  and (ii) taking into account the protostellar accretion that leads to the formation of the star on which then late accretion occurs. Indeed, regarding the latter point, previous studies rely on stellar models which start their evolution at an arbitrary point on the Hayashi track neglecting the protostellar accretion phase. 

This paper is a first attempt to address the effect of the accretion episode of helium rich gas on models consistently evolved starting from a small hydrostatic seed and then followed through the protostellar accretion phase. We analysed in detail the effect of the main physical parameters that affect the protostellar accretion phase and their impact on the following late accretion episode of polluted matter, with particular attention to the central helium abundance, the tracks position in the \hr{} diagram and the temporal evolution of surface lithium abundance.

Similarly to \citet{cassisi14} and \citet{dantona14}, we identified some weaknesses in the \citet{bastian13} scenario. It seems very unlikely to produce a discrete helium abundance pattern in the core that might reproduce the observed \ms{} splitting, without making strong assumptions on the amount of the accreted material (i.e. $M_\rmn{fin}~-~M_0$) and/or on the initial mass ($M_0$). 

Another crucial point is the amount of helium accreted into the stellar centre. As already shown by \citet{dantona14}, a consistent treatment of late accretion of polluted matter in stellar model computations shows that \pms{} stars develop a radiative core before the end of the accretion, thus preventing the further increase of central helium abundance during the accretion on \pms{} models. 

We proved that the age at which the radiative core forms depends not only on the initial mass ($M_0$) but also on the starting age of the late accretion $t_\rmn{0,l.acc}$. We found that adopting $t_\rmn{0,l.acc}$ values compatible with the \citet{bastian13} model, the later is the beginning of accretion of helium-rich matter and the earlier is the development of the radiative core, thus the lower the final central helium abundance when the final mass ($M_\rmn{fin}$) is reached. We showed that only starting the accretion at $t_\rmn{0,l.acc} \sim 0$ might lead to a relevant increase of the stellar central helium content. However, according to \citet{bastian13} for ages $\la 2$ Myr no helium rich polluted matter should yet be available.

The situation gets even worst if the protostellar accretion phase is included in stellar computations. We showed that in the \hr{} diagram the evolutionary tracks obtained with or without taking into account the protostellar accretion exhibit large differences during the \pms{} evolution but they tend to converge towards the \zams{} region. Clearly such morphological differences among the tracks cannot be directly tested in aglactic \gcs{} being confined in \pms{} phase. 

However, we showed that the inclusion of protostellar accretion has a drastic effect both on the central helium and on the surface lithium contents, even in \zams. Stellar models which take into account also the protostellar accretion phase from a very low mass seed (i.e. 0.001--0.050~\msun) have structures at the beginning of (or even during) the accretion of polluted matter significantly different from standard models which start their evolution from the top of the Hayashi track neglecting protostellar accretion.

As a general feature, models which follow protostellar accretion are less efficient in bringing polluted helium in the centre because of the faster growth of the radiative core, thus leading to lower central helium enhancement. The lower is the adopted $M_\rmn{seed}$ and the lower is the helium deposed into the stellar centre during the accretion phase because of the very quick development of a radiative core that, in the worst cases, forms before the end of the protostellar accretion phase itself. 

The mass accretion rate during the protostellar accretion plays a relevant role in determining the central helium and the survived surface lithium abundances, even in \zams. In all the cases the central helium content at the end of the accretion phase decreases if the accretion rate increases. We also showed that part of the helium accreted in the regions surrounding the centre can be dragged into the core at ages typical of the \zams{}, when a temporarily convective core forms. However, in all the cases, with the exception of models with $X_\rmn{d}~=~2\times10^{-5}$, the amount of central helium abundance after the late accretion processes is much lower than that generally derived to obtain the \ms{} splitting, while in the worst case the central helium abundance is not enhanced at all.

Another crucial point concerns the lithium abundance in accreting stars. Lithium in \gc{} turn-off and subgiant stars survives in the stellar mass range that we analysed in this paper. However, as already partially discussed in \citet{dantona14} and \citet{salaris14}, the inclusion of late accretion leads to a rapid and efficient lithium depletion during the \pms{}. We showed that if the protostellar accretion is accounted for, then models might fully deplete lithium during the first few million years, especially if the lowest $M_\rmn{seed}$ value and/or large accretion rate during the protostellar phase are adopted. This result is opposite to that predicted by standard non accreting models and, more importantly, to what obtained from the observations of surface lithium for \ms{} and turn-off stars in \gcs. 

\section*{Acknowledgements}
We would like to thank the reviewer Raffaele Gratton for carefully reading the manuscript and for his suggestions that helped us to improve the paper. This work has been supported by PRIN-MIUR 2010-2011 (chemical and dynamical evolution of the Milky Way and Local Group galaxies, PI: F. Matteucci), PRIN-INAF 2012 (the M4 core project with \emph{Hubble Space Telescope}, PI: L. Bedin), PRIN-INAF 2014 (the kaleidoscope of stellar populations in globular clusters with Hubble Space Telescope, PI: S. Cassisi) and by INFN (iniziativa specifica TAsP).

\bibliographystyle{mn2e}
\bibliography{bibliography}
\label{lastpage}

\begin{table*}
\centering
    \caption{Central helium and surface lithium and helium abundance at the end of the late accretion phase and at the age of $10^8$ yr for models with only late accretion (\dv). The logarithm of the effective temperature and luminosity for the $10^8$ yr model are also shown.}
	\label{tab:la_models}
\begin{tabular}{cccccccccc}
   \hline
   $M_0$ & $M_\rmn{fin}$ & \multicolumn{3}{l}{Late accretion end} & \multicolumn{5}{l}{$10^8$~yr model}\\
   (M$_{\sun})$ & (M$_\odot)$ & $^4$He(cent.) & $^4$He(surf.) & $^7$Li(surf.) & $^4$He(cent.) & $^4$He(surf.) & $^7$Li(surf.) & $\log~T_\rmn{eff}$ & $\log~L$/L$_{\sun}$ \\
   \hline
   \multicolumn{10}{c}{$t_\rmn{0,l.acc}~=$~0~Myr}\\
   \\  
   $0.200$ & $0.500$ & $0.330$ & $0.377$ & $ 5.145\times 10^{-16}$ & $0.338$ & $0.376$ & $ 1.569\times 10^{-16}$ & $3.677$ & $-0.970$ \\
   $0.200$ & $0.700$ & $0.338$ & $0.424$ & $ 3.244\times 10^{-12}$ & $0.353$ & $0.415$ & $ 3.162\times 10^{-12}$ & $3.784$ & $-0.225$ \\
   $0.400$ & $0.500$ & $0.271$ & $0.293$ & $ 4.868\times 10^{-10}$ & $0.276$ & $0.293$ & $ 1.465\times 10^{-10}$ & $3.640$ & $-1.150$ \\
   $0.400$ & $0.700$ & $0.286$ & $0.364$ & $ 6.516\times 10^{-10}$ & $0.298$ & $0.362$ & $ 6.475\times 10^{-10}$ & $3.760$ & $-0.362$ \\
   \\
   \multicolumn{10}{c}{$t_\rmn{0,l.acc}~=$~1~Myr}\\
   \\
   $0.200$ & $0.500$ & $0.324$ & $0.380$ & $ 1.400\times 10^{-24}$ & $0.334$ & $0.379$ & $ 2.245\times 10^{-25}$ & $3.677$ & $-0.972$ \\
   $0.200$ & $0.700$ & $0.330$ & $0.432$ & $ 2.390\times 10^{-16}$ & $0.347$ & $0.421$ & $ 2.321\times 10^{-16}$ & $3.785$ & $-0.227$ \\
   $0.400$ & $0.500$ & $0.268$ & $0.294$ & $ 2.805\times 10^{-10}$ & $0.273$ & $0.294$ & $ 8.355\times 10^{-11}$ & $3.640$ & $-1.151$ \\
   $0.400$ & $0.700$ & $0.280$ & $0.371$ & $ 4.629\times 10^{-10}$ & $0.292$ & $0.369$ & $ 4.600\times 10^{-10}$ & $3.761$ & $-0.366$ \\
   \\
   \multicolumn{10}{c}{$t_\rmn{0,l.acc}~=$~2~Myr}\\
   \\
   $0.200$ & $0.500$ & $0.318$ & $0.383$ & $ 0.0$ & $0.330$ & $0.382$ & $ 0.0$ & $3.677$ & $-0.975$ \\
   $0.200$ & $0.700$ & $0.326$ & $0.438$ & $ 4.976\times 10^{-31}$ & $0.341$ & $0.426$ & $ 4.822\times 10^{-31}$ & $3.786$ & $-0.229$ \\
   $0.400$ & $0.500$ & $0.266$ & $0.296$ & $ 1.262\times 10^{-10}$ & $0.271$ & $0.295$ & $ 3.766\times 10^{-11}$ & $3.640$ & $-1.151$ \\
   $0.400$ & $0.700$ & $0.274$ & $0.378$ & $ 2.873\times 10^{-10}$ & $0.288$ & $0.376$ & $ 2.855\times 10^{-10}$ & $3.762$ & $-0.369$ \\
   \\
   \multicolumn{10}{c}{$t_\rmn{0,l.acc}~=$~4~Myr}\\
   \\
   $0.200$ & $0.500$ & $0.310$ & $0.388$ & $ 0.0$ & $0.323$ & $0.387$ & $ 0.0$ & $3.677$ & $-0.978$ \\
   $0.200$ & $0.700$ & $0.335$ & $0.440$ & $ 0.0$ & $0.342$ & $0.428$ & $ 0.0$ & $3.786$ & $-0.230$ \\
   $0.400$ & $0.500$ & $0.260$ & $0.298$ & $ 1.603\times 10^{-11}$ & $0.266$ & $0.298$ & $ 4.756\times 10^{-12}$ & $3.640$ & $-1.154$ \\
   $0.400$ & $0.700$ & $0.265$ & $0.391$ & $ 6.014\times 10^{-11}$ & $0.280$ & $0.389$ & $ 5.976\times 10^{-11}$ & $3.763$ & $-0.373$ \\
	\hline
\end{tabular}
\end{table*}
\begin{table*}
\centering
\caption{Central helium and surface lithium and helium abundance at the end of the late accretion phase and at the age of $10^8$ yr for models with both protostellar and late accretion (\psa). The logarithm of the effective temperature and luminosity for the $10^8$ yr model are also shown.}
\label{tab:pla_models}
\begin{tabular}{cccccccccc}
   \hline
   $M_0$ & $M_\rmn{fin}$ & \multicolumn{3}{l}{Late accretion end} & \multicolumn{5}{l}{$10^8$~yr model}\\
   (M$_{\sun})$ & (M$_{\sun})$ & $^4$He(cent.) & $^4$He(surf.) & $^7$Li(surf.) & $^4$He(cent.) & $^4$He(surf.) & $^7$Li(surf.) & $\log~T_\rmn{eff}$ & $\log~L$/L$_{\sun}$ \\
   \hline
	\multicolumn{10}{c}{$M_\rmn{seed}~=~0.001~\rmn{M}_{\sun}$, $\dot{m}~=~10^{-5}$~M$_{\sun}$~yr$^{-1}$, $X_\rmn{d}~=~4\times10^{-5}$, $t_\rmn{0,l.acc}~=~2$~Myr}\\
\\ 
   $0.200$ & $0.500$ & $0.302$ & $0.392$ & $ 0.0$ & $0.316$ & $0.391$ & $ 0.0$ & $3.677$ & $-0.981$ \\
   $0.200$ & $0.700$ & $0.338$ & $0.439$ & $ 0.0$ & $0.345$ & $0.427$ & $ 0.0$ & $3.786$ & $-0.230$ \\
   $0.400$ & $0.500$ & $0.252$ & $0.309$ & $ 1.115\times 10^{-20}$ & $0.253$ & $0.309$ & $ 3.255\times 10^{-21}$ & $3.640$ & $-1.158$ \\
   $0.400$ & $0.700$ & $0.252$ & $0.428$ & $ 5.193\times 10^{-22}$ & $0.261$ & $0.425$ & $ 5.153\times 10^{-22}$ & $3.766$ & $-0.384$ \\
   \\
   \multicolumn{10}{c}{$M_\rmn{seed}~=~0.010~\rmn{M}_{\sun}$, $\dot{m}~=~10^{-5}$~M$_{\sun}$~yr$^{-1}$, $X_\rmn{d}~=~4\times10^{-5}$, $t_\rmn{0,l.acc}~=~2$~Myr}\\
   \\
   $0.200$ & $0.500$ & $0.311$ & $0.387$ & $ 0.0$ & $0.324$ & $0.386$ & $ 0.0$ & $3.677$ & $-0.977$ \\
   $0.200$ & $0.700$ & $0.334$ & $0.440$ & $ 0.0$ & $0.341$ & $0.427$ & $ 0.0$ & $3.786$ & $-0.230$ \\
   $0.400$ & $0.500$ & $0.261$ & $0.297$ & $ 2.902\times 10^{-11}$ & $0.267$ & $0.297$ & $ 8.669\times 10^{-12}$ & $3.640$ & $-1.153$ \\
   $0.400$ & $0.700$ & $0.267$ & $0.389$ & $ 9.381\times 10^{-11}$ & $0.282$ & $0.386$ & $ 9.322\times 10^{-11}$ & $3.762$ & $-0.372$ \\
   \\
   \multicolumn{10}{c}{$M_\rmn{seed}~=~0.050~\rmn{M}_{\sun}$, $\dot{m}~=~10^{-5}$~M$_{\sun}$~yr$^{-1}$, $X_\rmn{d}~=~4\times10^{-5}$, $t_\rmn{0,l.acc}~=~2$~Myr}\\
   \\
   $0.200$ & $0.500$ & $0.313$ & $0.386$ & $ 0.0$ & $0.326$ & $0.385$ & $ 0.0$ & $3.677$ & $-0.976$ \\
   $0.200$ & $0.700$ & $0.332$ & $0.440$ & $ 0.0$ & $0.340$ & $0.427$ & $ 0.0$ & $3.786$ & $-0.230$ \\
   $0.400$ & $0.500$ & $0.263$ & $0.297$ & $ 5.677\times 10^{-11}$ & $0.268$ & $0.296$ & $ 1.738\times 10^{-11}$ & $3.640$ & $-1.152$ \\
   $0.400$ & $0.700$ & $0.270$ & $0.384$ & $ 1.679\times 10^{-10}$ & $0.284$ & $0.382$ & $ 1.669\times 10^{-10}$ & $3.762$ & $-0.371$ \\
   \\
   \multicolumn{10}{c}{$M_\rmn{seed}~=~0.001~\rmn{M}_{\sun}$, $\dot{m}~=~10^{-6}$~M$_{\sun}$~yr$^{-1}$, $X_\rmn{d}~=~4\times10^{-5}$, $t_\rmn{0,l.acc}~=~2$~Myr}\\
   \\
   $0.200$ & $0.500$ & $0.310$ & $0.388$ & $ 0.0$ & $0.323$ & $0.387$ & $ 0.0$ & $3.677$ & $-0.977$ \\
   $0.200$ & $0.700$ & $0.335$ & $0.440$ & $ 0.0$ & $0.342$ & $0.428$ & $ 0.0$ & $3.786$ & $-0.230$ \\
   $0.400$ & $0.500$ & $0.258$ & $0.300$ & $ 3.447\times 10^{-12}$ & $0.263$ & $0.299$ & $ 1.041\times 10^{-12}$ & $3.640$ & $-1.154$ \\
   $0.400$ & $0.700$ & $0.260$ & $0.398$ & $ 1.560\times 10^{-11}$ & $0.277$ & $0.395$ & $ 1.550\times 10^{-11}$ & $3.763$ & $-0.375$ \\
   \\
   \multicolumn{10}{c}{$M_\rmn{seed}~=~0.001~\rmn{M}_{\sun}$, $\dot{m}~=~10^{-4}$~M$_{\sun}$~yr$^{-1}$, $X_\rmn{d}~=~4\times10^{-5}$, $t_\rmn{0,l.acc}~=~2$~Myr}\\
   \\
   $0.200$ & $0.500$ & $0.323$ & $0.397$ & $ 0.0$ & $0.326$ & $0.396$ & $ 0.0$ & $3.678$ & $-0.982$ \\
   $0.200$ & $0.700$ & $0.347$ & $0.433$ & $ 0.0$ & $0.353$ & $0.421$ & $ 0.0$ & $3.786$ & $-0.228$ \\
   $0.400$ & $0.500$ & $0.252$ & $0.326$ & $ 0.0$ & $0.253$ & $0.326$ & $ 0.0$ & $3.640$ & $-1.163$ \\
   $0.400$ & $0.700$ & $0.257$ & $0.439$ & $ 0.0$ & $0.261$ & $0.436$ & $ 0.0$ & $3.767$ & $-0.390$ \\
   \\
   \multicolumn{10}{c}{$M_\rmn{seed}~=~0.001~\rmn{M}_{\sun}$, $\dot{m}~=~10^{-5}$~M$_{\sun}$~yr$^{-1}$, $X_\rmn{d}~=~2\times10^{-5}$, $t_\rmn{0,l.acc}~=~2$~Myr}\\
   \\
   $0.200$ & $0.500$ & $0.365$ & $0.365$ & $ 0.0$ & $0.367$ & $0.364$ & $ 0.0$ & $3.676$ & $-0.959$ \\
   $0.200$ & $0.700$ & $0.378$ & $0.407$ & $ 0.0$ & $0.385$ & $0.400$ & $ 0.0$ & $3.782$ & $-0.216$ \\
   $0.400$ & $0.500$ & $0.290$ & $0.290$ & $ 0.0$ & $0.291$ & $0.289$ & $ 0.0$ & $3.639$ & $-1.138$ \\
   $0.400$ & $0.700$ & $0.314$ & $0.377$ & $ 0.0$ & $0.319$ & $0.375$ & $ 0.0$ & $3.761$ & $-0.360$ \\
	\\
   \multicolumn{10}{c}{$M_\rmn{seed}~=~0.001~\rmn{M}_{\sun}$, $\dot{m}~=~10^{-5}$~M$_{\sun}$~yr$^{-1}$, $X_\rmn{d}~=~4\times10^{-5}$, $t_\rmn{0,l.acc}~=~0$~Myr}\\
   \\
   $0.200$ & $0.500$ & $0.307$ & $0.389$ & $ 0.0$ & $0.321$ & $0.389$ & $ 0.0$ & $3.677$ & $-0.978$ \\
   $0.200$ & $0.700$ & $0.337$ & $0.440$ & $ 0.0$ & $0.344$ & $0.429$ & $ 0.0$ & $3.786$ & $-0.230$ \\
   $0.400$ & $0.500$ & $0.252$ & $0.307$ & $ 9.776\times 10^{-18}$ & $0.254$ & $0.307$ & $ 3.019\times 10^{-18}$ & $3.640$ & $-1.157$ \\
   $0.400$ & $0.700$ & $0.252$ & $0.423$ & $ 5.956\times 10^{-18}$ & $0.264$ & $0.420$ & $ 5.910\times 10^{-18}$ & $3.765$ & $-0.382$ \\
   \\
   \multicolumn{10}{c}{$M_\rmn{seed}~=~0.001~\rmn{M}_{\sun}$, $\dot{m}~=~10^{-5}$~M$_{\sun}$~yr$^{-1}$, $X_\rmn{d}~=~4\times10^{-5}$, $t_\rmn{0,l.acc}~=~4$~Myr}\\
   \\
   $0.200$ & $0.500$ & $0.306$ & $0.394$ & $ 0.0$ & $0.313$ & $0.393$ & $ 0.0$ & $3.677$ & $-0.982$ \\
   $0.200$ & $0.700$ & $0.339$ & $0.438$ & $ 0.0$ & $0.345$ & $0.428$ & $ 0.0$ & $3.787$ & $-0.230$ \\
   $0.400$ & $0.500$ & $0.252$ & $0.311$ & $ 7.658\times 10^{-24}$ & $0.253$ & $0.311$ & $ 2.323\times 10^{-24}$ & $3.640$ & $-1.160$ \\
   $0.400$ & $0.700$ & $0.252$ & $0.432$ & $ 1.467\times 10^{-26}$ & $0.259$ & $0.429$ & $ 1.456\times 10^{-26}$ & $3.766$ & $-0.385$ \\
	\hline
\end{tabular}
\end{table*}

\end{document}